\documentclass[a4paper,11pt]{article}

\usepackage{jheppub}
\usepackage{graphicx}
\usepackage{amssymb}
\usepackage{amsmath}
\usepackage{hyperref}

\DeclareMathOperator{\tr}{tr}
\DeclareMathOperator{\adj}{adj}
\DeclareMathOperator{\rank}{rank}
\DeclareMathOperator{\diag}{diag}
\newcommand{\abs}[1]{\lvert #1 \rvert}

\usepackage[table]{xcolor}
\definecolor{lightgray}{gray}{0.9}

\title{\boldmath Is our vacuum global in a 331 model with three triplets?}

\author{Kristjan Kannike,} 

\author{Niko Koivunen,}

\author[1]{Aleksei Kubarski\note{Corresponding author.}}

\affiliation{National Institute of Chemical Physics and Biophysics, \\ R\"{a}vala 10, 10143 Tallinn, Estonia}

\emailAdd{Kristjan.Kannike@cern.ch}

\emailAdd{Niko.Koivunen@kbfi.ee} 

\emailAdd{Aleksei.Kubarski@ut.ee} 

\abstract{We consider a 331 model, based on $\beta=-1/\sqrt{3}$, with three $SU(3)$ triplets with a softly broken $\mathbb{Z}_2$ symmetry.  The resulting scalar potential is commonly used in phenomenology. We systematically determine all the potential minima and obtain the conditions under which the electroweak vacuum is global with the help of orbit space methods. For the case the electroweak vacuum is not global, we calculate bounds on the scalar couplings from metastability. We find a parametrisation of the potential couplings in terms of physical quantities and use it to show the available parameter space.}

\begin{document}
\maketitle
\flushbottom

%%%%%%%%%%%%%%%%%%%%%%%%%%%%%%%%%%%%%%%%%%%%%%%%%%%%%%%%%%
\section{Introduction}
\label{sec:intro}

The Standard Model (SM) still harbours unanswered questions: for example, why are there exactly three generations of matter fields? An appealing answer is given by 331-models which are based on the $SU(3)_c\times SU(3)_L \times U(1)_X$ gauge symmetry. These models have the ability to explain the number of fermion families in nature, since the cancellation of gauge anomalies is different from the SM.
The $SU(3)_c\times SU(3)_L\times U(1)_X$ gauge group has one additional diagonal generator compared to the SM. Therefore the 331-models have freedom in the definition of the electric charge,
\begin{equation}\label{electric charge}
Q=T_3+\beta T_8+X,
\end{equation}
where the $T_3$ and $T_8$ are the diagonal $SU(3)_L$ generators, $X$ is the $U_X$-charge and the parameter $\beta$ can have any value. The most studied models, corresponding to $\beta=\pm 1/\sqrt{3}$ \cite{Georgi:1978bv,Singer:1980sw, Valle:1983dk, Montero:1992jk, Foot:1994ym, Long:1995ctv, Long:1996rfd, Pleitez:1994pu, Long:1997vbr,Dong:2013ioa} and $\beta=\pm \sqrt{3}$ \cite{Pisano:1992bxx, Frampton:1992wt, Foot:1992rh, Tonasse:1996cx, Nguyen:1998ui},  differ significantly in their particle content.
The models based on  $\beta=\pm\sqrt{3}$ contain particles with exotic electric charges, such as new doubly charged scalars and gauge bosons, and new quarks with electric charges $\pm 4/3$ and $\pm 5/3$. The models based on $\beta=\pm1/\sqrt{3}$, on the other hand, do not contain such particles.

The 331 models have complicated scalar sectors. 
The spontaneous symmetry breaking of $SU(3)_L\times U(1)_X\to U(1)_{\rm EM}$ requires only two scalar triplets, resulting in a relatively simple scalar potential, with the caveat that some of the fermions will be massless at tree level. In the case of $\beta=\pm 1/\sqrt{3}$, radiative corrections are required in order to generate all fermion masses \cite{Ponce:2002sg,Dong:2006mg}. For models with $\beta=\pm\sqrt{3}$, the situation is even more grim as effective operators are needed to generate all the fermion masses \cite{Ferreira:2011hm, Cogollo:2013mga, Dong:2014esa}.  
Models of both types need three scalar triplets in order to generate tree-level masses to all the particles. The models with  $\beta=\pm\sqrt{3}$ also require an additional scalar sextet in order to give tree-level masses for all of the charged leptons, making the general scalar potential extremely complicated \cite{DeConto:2015eia}. The potential in  $\beta=\pm 1/\sqrt{3}$ models, on the other hand, is complicated by  the requirement of two scalar triplets to be in the same representation, which produces multiple cross terms. A systematic study of the scalar sector of 331 models for different values of $\beta$ has been conducted in \cite{Diaz:2003dk}. They show that the 331-model reduces to the two-Higgs-doublet model (2HDM) in the decoupling limit. Constraints for the 331 scalar potential for a general value of $\beta$, such as boundedness from below, are studied in \cite{Costantini:2020xrn}.

The scalar potential of 331 models has not been examined in the same depth as, for example, in the 2HDM. The existence of multiple local minima has not been studied much. 
To the best of our knowledge there have been only three papers where the global minimum of the 331 scalar potential has been considered \cite{Giraldo:2009yi,Giraldo:2011gd,Dorsch:2024ddk}. The first two investigate the potential in $\beta=\pm 1/\sqrt{3}$ model,  using methods originally developed for the 2HDM \cite{Maniatis:2006fs,Ivanov:2006yq,Ivanov:2007de}. The first paper studies a potential with two scalar triplets  \cite{Giraldo:2009yi}. The second one studies a potential with three scalar triplets without a trilinear $f$-term \cite{Giraldo:2011gd}, which is often included in order to avoid Goldstones in the physical spectrum.\footnote{The appearance of a physical Goldstone is avoided in \cite{Giraldo:2011gd} by demanding a specific relation between vacuum expectation values, which guarantees that the accidental continuous symmetry remains unbroken.} The appearance of a global radiatively induced minimum at high scales in the one-loop effective potential was studied in \cite{Dorsch:2024ddk}.

In present work we will study the $\beta=-1/\sqrt{3}$ model with three scalar triplets. 
Many variants of the scalar potential occur in the literature, due to the fact that the most general scalar potential in such models is quite complicated \cite{Huitu:2024nap}, because there are two scalar triplets in the same representation, allowing for multiple cross terms. 
In phenomenological studies the number of terms is often reduced through the use of either discrete or continuous symmetries. For the former case, the most common is a $\mathbb{Z}_{2}$ symmetry, which often also limits the Yukawa interactions as well \cite{Pal:1994ba,Sanchez-Vega:2016dwe,Pinheiro:2022bcs}. The continuous symmetries are often $U(1)$-symmetries, associated to the lepton number, or flavour symmetries \cite{Pinheiro:2022bcs,Huitu:2017ukq}. A simple potential has been used, for example, to study collider phenomenology \cite{Cao:2016uur, Alves:2022hcp}, flavour physics \cite{Oliveira:2022vjo,Buras:2012dp} and dark matter \cite{Mizukoshi:2010ky,Ruiz-Alvarez:2012nvg, Profumo:2013sca}. 

We study a scalar potential that includes all the terms of the simple potential used in the phenomenological studies \cite{Sanchez-Vega:2018qje}. Our main goal is to study the (meta)stability of our electroweak vacuum. We work out the complete structure of the extrema of the scalar potential and conditions for their appearance. To that end, we determine the orbit space of the three triplets with $P$-matrix methods \cite{Abud:1983id,Abud:1981tf,Sartori:2005sh,Talamini:2006wd}. Using the orbit space, we reduce the large number of real field degrees of freedom to a small number of orbit variables which considerably simplifies finding the potential minima \cite{Kim:1981xu,Degee:2012sk,Heikinheimo:2017nth}.  

First of all, the scalar potential needs to be bounded from below: we complete the necessary bounded-from-below conditions given in \cite{Sanchez-Vega:2018qje}. Then we study the parameter space where our vacuum is the global one. But even when the electroweak vacuum is not global, it can be metastable with a lifetime longer than the age of the Universe. We thus study tunnelling from ours into other vacua and determine bounds on the parameter space from metastability with the FindBounce code \cite{Guada:2018jek,Guada:2020xnz} to calculate tunnelling rates. We illustrate the results for a typical parameter space in the limit of large $v_{\chi} \approx f$ and all heavy masses, except one, equal to a common $m_{331}$ mass scale. We find that the electroweak vacuum may not be global if the mixing of the $\eta$ and $\rho$ triplets with the triplet $\chi$ is non-zero. Still a large part of this parameter space is metastable. On the other  hand, in the part of the typical parameter space where this mixing is negligible, the model is unitary, the potential bounded from below and the electroweak vacuum is global.

This article is structured as follows. In section~\ref{sec:model}, we review the  331 model with three triplets. In section~\ref{orbit:space}, we define the orbit space of the model and determine its shape. Section~\ref{sec:exp:constr} gives the main constraints: unitarity, boundedness-from-below of the scalar potential, and metastability of the electroweak vacuum. In section~\ref{sec:EW:vacuum}, we study the mass spectrum in our vacuum and parametrise the scalar couplings via physical quantities. We  study the (meta)stability of our vacuum in section~\ref{sec:stab}. Our conclusions are given in section~\ref{sec:concl}.

%%%%%%%%%%%%%%%%%%%%%%%%%%%%%%%%%%%%%%%%%%%%%%%%%%%%%%%%%%
\section{331 model with three triplets}
\label{sec:model}
We concentrate on the scalar sector of a 331 model where the electric charge, eq.~\eqref{electric charge} is determined by $\beta=-1/\sqrt{3}$. Because we focus on the scalar sector, we do not discuss the fermion and gauge sector: the full particle content can be found in ref.~\cite{Huitu:2024nap}, for example.\footnote{The quark sector in 331 models always has tree-level FCNCs. The suppression of these FCNCs is determined by the quark diagonalisation matrices and the masses of the BSM particles. These, in turn, are controlled by the quark Yukawa couplings and the $SU(3)_L$-breaking VEV $v_\chi$. Because we work at tree-level, the fermion sector does not directly influence the structure of the scalar potential. In the numerical examples we study, we have $v_\chi = 10$ TeV, which is enough to suppress the FCNCs, assuming Yukawa couplings with reasonable structure, as discussed for example in \cite{Huitu:2024nap}.} 
There are three scalar triplets, given by
\begin{equation}
  \rho = 
  \begin{pmatrix}
    \rho_{1}^{+}
    \\
    \rho_{2}^{0}
    \\
    \rho_{3}^{+}
  \end{pmatrix}
  \sim (1, \mathbf{3}, 2/3),
  \quad
  \eta =
  \begin{pmatrix}
    \eta_{1}^{0}
    \\
    \eta_{2}^{-}
    \\
    \eta_{3}^{0}
  \end{pmatrix}
  \sim (1, \mathbf{3}, -1/3),
  \quad
  \chi = 
  \begin{pmatrix}
    \chi_{1}^{0}
    \\
    \chi_{2}^{-}
    \\
    \chi_{3}^{0}
  \end{pmatrix}
  \sim (1, \mathbf{3}, -1/3).
\label{eq:triplets}
\end{equation}

We study the scalar potential invariant under the 331 group and the $\mathbb{Z}_{2}$ symmetry under which $\chi \to -\chi$ \cite{Foot:1992rh,Foot:1994ym,Dias:2005yh,Dong:2006gx,Mizukoshi:2010ky,Montero:2011tg,Cogollo:2014jia,Sanchez-Vega:2016dwe}, broken only softly, given by
\begin{equation}
\label{eq:V}
\begin{split}
  V &= \mu_{\eta}^{2} \eta^{\dagger} \eta + \mu_{\rho}^{2} \rho^{\dagger} \rho + \mu_{\chi}^{2} \chi^{\dagger} \chi
  + \lambda_{\eta} (\eta^{\dagger} \eta)^{2} + \lambda_{\rho} (\rho^{\dagger} \rho)^{2} + \lambda_{\chi} (\chi^{\dagger} \chi)^{2} 
  \\
  & + \lambda_{\eta\rho} (\eta^{\dagger} \eta) (\rho^{\dagger} \rho) + \lambda_{\eta \chi} (\eta^{\dagger} \eta) (\chi^{\dagger} \chi)
  +\lambda_{\rho\chi} (\rho^{\dagger} \rho) (\chi^{\dagger} \chi)
  \\
  & + \lambda'_{\eta\rho} (\eta^{\dagger} \rho ) (\rho^{\dagger} \eta) + \lambda'_{\eta \chi} (\eta^{\dagger} \chi ) (\chi^{\dagger} \eta)
  +\lambda'_{\rho\chi} (\rho^{\dagger} \chi) (\chi^{\dagger} \rho)
  \\
  & + \frac{1}{2} \lambda''_{\eta\chi} (\chi^{\dagger} \eta)^{2} - \frac{f}{\sqrt{2}} \epsilon^{ijk} \eta_{i} \rho_{j} \chi_{k} + \text{h.c.},
\end{split}
\end{equation}
where we take $f > 0$ and $\lambda''_{\eta\chi} < 0$ without loss of generality.\footnote{The trilinear $f$ term softly breaks the $\mathbb{Z}_{2}$, which is necessary to avoid the appearance of an axion that is ruled out \cite{Bardeen:1986yb,Carvajal:2017gjj}. As usual, we do not consider the soft-breaking mass term $\chi^{\dagger} \eta$.} This is the same potential as studied in ref. \cite{Sanchez-Vega:2018qje} and it closely resembles the potential used in the phenomenological studies \cite{Cao:2016uur, Alves:2022hcp,Oliveira:2022vjo,Buras:2012dp,Mizukoshi:2010ky,Ruiz-Alvarez:2012nvg, Profumo:2013sca}, with a $\lambda''_{\eta\chi}$ additional term.\footnote{The scalar couplings are related to those of ref.~\cite{Sanchez-Vega:2018qje} as: $\mu_{1}^{2} = -\mu_{\eta}^{2}$, $\mu_{2}^{2} = -\mu_{\rho}^{2}$, $\mu_{3}^{2} = -\mu_{\chi}^{2}$, $\lambda_{1} = \lambda_{\eta}$, $\lambda_{2} = \lambda_{\rho}$, $\lambda_{3} = \lambda_{\chi}$, $\lambda_{4} = \lambda_{\eta\chi}$, $\lambda_{5} = \lambda_{\rho\chi}$, $\lambda_{6} = \lambda_{\eta\rho}$, $\lambda_{7} = \lambda'_{\eta\chi}$, $\lambda_{8} = \lambda'_{\rho\chi}$, $\lambda_{9} = \lambda'_{\eta\rho}$, $\lambda_{10} = \lambda''_{\eta\chi}/2$.} Notice that in the trilinear term, the invariant $\epsilon^{ijk} \eta_{i} \rho_{j} \chi_{k} = \det (\eta \rho \chi)$, the determinant of the matrix whose column vectors are $\eta$, $\rho$ and $\chi$.

%%%%%%%%%%%%%%%%%%%%%%%%%%%%%%%%%%%%%%%%%%%%%%%%%%%%%%%%%%
\section{Orbit space}
\label{orbit:space}

%%%%%%%%%%%%%%%%%%%%%%%%%%%%%%%%%%%%%%%%%%%%%%%%%%%%%%%%%%
\subsection{Orbit space and scalar potential}
\label{orbit:space:potential}

The scalar potential is invariant under the symmetry group $G$ of the theory. Although the potential generally depends on a large number of real fields, it does so in terms of a smaller number of gauge invariants that constitute the orbit space of the theory. The orbit of a constant field configuration $\phi$ (such as a VEV) is the set of states $\phi_{\theta} = T(\theta) \phi$ with $T(\theta)$ an element of the group $G$. All the $\phi_{\theta}$ states respect the same group, the little group of the orbit, as does $\phi$. If the group $G$ is unitary then all the states $\phi_{\theta}$ have the same norm $\phi^{\dagger} \phi$. Furthermore, it is often useful to separate the orbit space into non-negative radial, dimensionful, field norms and finite, dimensionless, orbit variables. While the fields rotate under gauge transformations, the invariants do not change: a gauge orbit in field space is shrunk to a point in orbit space. In particular, the value of the scalar potential is left unchanged. It is therefore convenient to go from field space to orbit space to study vacuum stability and minima of the potential. The tradeoff is in that the shape of the orbit space is non-trivial. To find the equations and inequalities that define the orbit space, we can define an invariant polynomial basis and use the $P$-matrix formalism \cite{Abud:1983id,Abud:1981tf,Talamini:2006wd} (see \cite{Sartori:2005sh} for an overview).

Sometimes (higher-order) invariants that do not enter the potential are needed to express all $P$-matrix elements. Then we need to extend our initial guess for the basis. The simplest way to do that is to promote the $P$-matrix element that could not be expressed to a new basis element, and repeat the procedure until all $P$-matrix elements can be expressed in the extended basis. 

The basis of the orbit space of the 331 model is given by the invariants of the $SU(3)_c \times SU(3)_L$ gauge group:
\begin{equation}
\begin{aligned}
  p_{1} &= \eta^{\dagger} \eta, 
  & 
  p_{2} &= \rho^{\dagger} \rho, 
  & 
  p_{3} &= \chi^{\dagger} \chi,
  \\
  p_{4} &= \Re \chi^{\dagger} \eta, 
  & 
  p_{5} &= \Re  \rho^{\dagger} \chi, 
  & 
  p_{6} &= \Re \eta^{\dagger} \rho,
  \\
  p_{7} &= \Re \det (\eta \rho \chi), 
  & 
  p_{8} &=  \Im \chi^{\dagger} \eta, 
  &
  p_{9} &= \Im \rho^{\dagger} \chi,
  \\
  p_{10} &= \Im \eta^{\dagger} \rho,
  &
  p_{11} &= \Im \det (\eta \rho \chi).
\end{aligned}
\label{eq:basis}
\end{equation}
A $U(1)$ factor of the full gauge group, such as the $U(1)_{X}$ subgroup, does not contribute to the non-trivial structure of the orbit space because it has only one-dimensional representations. We want the basis to be as simple and small as possible, so that the shape of the orbit space could be found more easily. This is the reason that we use the basis \eqref{eq:basis} invariant only under the $SU(3)_c \times SU(3)_L$ gauge group and not under $U(1)_{X}$. Imposing the $U(1)_{X}$ subgroup on the basis would forbid some triplet bilinears and force us to use their Hermitian squares as polynomial basis elements: instead of $\chi^{\dagger} \rho$, for example, we would have to use $(\chi^{\dagger} \rho) (\rho^{\dagger} \chi)$. Thus, imposing $U(1)_{X}$ on the basis \eqref{eq:basis} would forbid the lower-dimensional $p_{5}$, $p_{6}$, $p_{9}$ and $p_{10}$, and we would have to use their squares in the basis. This, in turn, would strongly complicate finding the shape of the orbit space. For this reason, we only impose the full gauge symmetry on the scalar potential \eqref{eq:V}: although the basis contains elements not invariant under the $U(1)$, only fully gauge-symmetric combinations of the basis elements appear in the potential. The potential, in terms of the invariants, is given by
\begin{equation}
\label{eq:V:inv:real}
\begin{split}
  V &= \mu_{\eta}^{2} p_{1} + \mu_{\rho}^{2} p_{2} + \mu_{\chi}^{2} p_{3}
  + \lambda_{\eta} p_{1}^{2} + \lambda_{\rho} p_{2}^{2} + \lambda_{\chi} p_{3}^{2} + \lambda_{\eta\rho} p_{1} p_{2} 
  + \lambda_{\eta \chi} p_{1} p_{3} \\
  & 
    + \lambda_{\rho\chi} p_{2} p_{3}  + \lambda'_{\eta \chi} (p_{4}^{2} + p_{8}^{2}) - \abs{\lambda''_{\eta\chi}} (p_{4}^{2} - p_{8}^{2})
   + \lambda'_{\rho\chi} (p_{5}^{2} + p_{9}^{2}) 
  \\
  &\lambda'_{\eta\rho} (p_{6}^{2} + p_{10}^{2}) 
  - \sqrt{2} \abs{f} \, p_{7},
\end{split}
\end{equation}
where we have set $\lambda''_{\eta\chi} < 0$ and $f > 0$ and $p_{11} = 0$ by phase rotations.

In order to separate the orbit space into radial field norms and angular variables, we hence define the dimensionless orbit space variables
\begin{equation}
\label{eq:orbit:variables}
  \vartheta_{1}^{2} = \frac{p_{4}^{2} + p_{8}^{2}}{p_{1} p_{3}}, 
  \qquad
  \vartheta_{2}^{2} = \frac{p_{5}^{2} + p_{9}^{2}}{p_{2} p_{3}}, 
  \qquad
  \vartheta_{3}^{2} = \frac{p_{6}^{2} + p_{10}^{2}}{p_{1} p_{2}}, 
  \qquad
  \vartheta_{4} = \frac{p_{7}}{\sqrt{p_{1} p_{2} p_{3}}}.
\end{equation}
Henceforth we will usually mean by orbit space the space of these variables.
The correspondence between our notation and that of ref.~\cite{Sanchez-Vega:2018qje} is given by $\theta_{1} = \vartheta_{1}^{2}$, $\theta_{2} = \vartheta_{2}^{2}$, $\theta_{3} = \vartheta_{3}^{2}$, $\theta_{4} = 2 \vartheta_4$. Collectively, we also refer to the `spatial' variables $\vartheta_{1}$, $\vartheta_{2}$ and $\vartheta_{3}$ as $\vartheta_{i}$. It is easy to see, using the Cauchy–Schwarz inequality, that the orbit space lies within the hypercube
\begin{equation}
  -1 \leq \vartheta_{i}, \vartheta_{4} \leq 1,
\end{equation}
but its actual shape is non-trivial. Only if one of the triplets is zero, the orbit space is an interval: if e.g. $\rho = 0$, then the orbit space is given by the interval $\vartheta_{1} \in [-1, 1]$.

There is a simple geometric interpretation of the boundary of the orbit space. For real field values, the orbit variables are given by cosines of the angles between triplets: $\vartheta_{1} = \cos (\angle \chi \eta)$, $\vartheta_{2} = \cos (\angle \chi \rho)$ and $\vartheta_{3} = \cos (\angle \eta \rho)$. Because $\epsilon^{ijk} \eta_{i} \rho_{j} \chi_{k} = \det (\eta \rho \chi)$ is the signed volume of the parallelepiped generated by real $\eta$, $\rho$ and $\chi$, the orbit variable $\vartheta_{4}$ is the signed volume of the parallelepiped generated by the respective unit vectors.

The geometric interpretation leads us to discover a \emph{syzygy}: a relation between the basis elements. Generalising the argument used to prove the parallelepiped volume formula \eqref{eq:cell} to complex vectors, we have
\begin{equation}
\begin{split}
  \abs{\det (\eta \rho \chi)}^{2} 
  &= \det (\eta \rho \chi)^{\dagger} \det (\eta \rho \chi)
  = \det (\eta \rho \chi)^{\dagger} (\eta \rho \chi)
  = \det
  \begin{pmatrix}
    \eta^{\dagger} \eta & \eta^{\dagger} \rho & \eta^{\dagger} \chi
    \\
    \rho^{\dagger} \eta & \rho^{\dagger} \rho & \rho^{\dagger} \chi
    \\
    \chi^{\dagger} \eta & \chi^{\dagger} \rho & \chi^{\dagger} \chi
  \end{pmatrix}
  \\
  &= \abs{\eta}^{2} \abs{\rho}^{2} \abs{\chi}^{2} - \abs{\rho}^{2} (\chi^{\dagger} \eta) (\eta^{\dagger} \chi)
  -\abs{\eta}^{2} (\chi^{\dagger} \rho) (\rho^{\dagger} \chi) - \abs{\chi}^{2} (\eta^{\dagger} \rho) (\rho^{\dagger} \eta)
  \\
  &+ (\chi^{\dagger} \eta) (\eta^{\dagger} \rho) (\rho^{\dagger} \chi) + (\eta^{\dagger} \chi) (\chi^{\dagger} \rho) (\rho^{\dagger} \eta)
\end{split}
\end{equation}
or
\begin{equation}
\begin{split}
  p_{7}^{2} + p_{11}^{2} &= p_{1} p_{2} p_{3} - p_{2} (p_{4}^{2} + p_{8}^{2}) - p_{1} (p_{5}^{2} + p_{9}^{2})
  - p_{3} (p_{6}^{2} + p_{10}^{2}) \\
  &+ 2 \Re (p_{4} + i p_{8}) (p_{6} + i p_{10}) (p_{5} + i p_{9}). 
\end{split}
\label{eq:syzygy}
\end{equation}
The orbit space lies in the semialgebraic set defined by the syzygy. 

In terms of the orbit variables and field norms, the scalar potential is, at a minimum (see below), given by
\begin{equation}
\label{eq:V:orbit:real}
\begin{split}
  V &= \mu_{\eta}^{2} \abs{\eta}^{2} + \mu_{\rho}^{2} \abs{\rho}^{2} + \mu_{\chi}^{2} \abs{\chi}^{2}
  + \lambda_{\eta} \abs{\eta}^{4} + \lambda_{\rho} \abs{\rho}^{4} + \lambda_{\chi} \abs{\chi}^{4}
  \\
  & 
  + [\lambda_{\eta \chi} + (\lambda'_{\eta \chi} - \abs{\lambda''_{\eta\chi}}) \vartheta_{1}^{2}] \abs{\eta}^{2} \abs{\chi}^{2}
  + (\lambda_{\rho\chi} + \lambda'_{\rho\chi} \vartheta_{2}^{2}) \abs{\rho}^{2}  \abs{\chi}^{2} 
  \\
  &+ (\lambda_{\eta\rho} + \lambda'_{\eta\rho} \vartheta_{3}^{2}) \abs{\eta}^{2} \abs{\rho}^{2} 
  - \sqrt{2} \abs{f} \, \vartheta_{4} \, \abs{\eta} \abs{\rho} \abs{\chi}.
\end{split}
\end{equation}

%%%%%%%%%%%%%%%%%%%%%%%%%%%%%%%%%%%%%%%%%%%%%%%%%%%%%%%%%%
\subsection{Shape of the orbit space}
\label{orbit:space:shape}

The shape of the orbit space is found with the help of the $P$-matrix formalism \cite{Abud:1983id,Abud:1981tf,Sartori:2005sh,Talamini:2006wd}. The $P$-matrix is defined by
\begin{equation}
  P_{ij} = \frac{\partial p_{i}}{\partial \Phi^{\dagger}_{a}} \frac{\partial p_{j}}{\partial \Phi^{a}},
\label{eq:p:matrix:def}
\end{equation}
where $\Phi_{a}$ runs over all the field components: it is the Hermitian square of the Jacobian matrix of the polynomial basis. The elements of the $P$-matrix are gauge invariants themselves, which means that they can be expressed in terms of the polynomial basis elements. 

We have computed the $P$-matrix specified by eq. \eqref{eq:p:matrix:def} with the Wolfram Mathematica computer algebra system, using the whole polynomial basis \eqref{eq:basis}. It is easy to show that the invariants $p_{8}$, $p_{9}$, $p_{10}$ and $p_{11}$ --- the imaginary parts of the field bilinears and the trilinear --- are non-zero only in the interior of the orbit space (since adding phases drives the orbit variables away from extremal values). Because of that, we set these invariants to zero when we calculate the orbit space boundary. In particular, if we consider real fields, we immediately have $p_{8} = p_{9} = p_{10} = p_{11} = 0$. We will see that all potential minima do correspond to real field configurations.
The $P$-matrix for the relevant invariants is then given by
\begin{equation}
  P = 
  \begin{pmatrix}
    2 p_{1} & 0 & 0 & p_{4} & 0 & p_{6} & p_{7}
    \\
    0 & 2 p_{2} & 0 & 0 & p_{5} & p_{6} & p_{7}
    \\
    0 & 0 & 2 p_{3} & p_{4} & p_{5} & 0 & p_{7}
    \\
    p_{4} & 0 & p_{4} & \frac{p_{1} + p_{3}}{2} & \frac{p_{6}}{2} & \frac{p_{5}}{2} & 0
    \\
    0 & p_{5} & p_{5} & \frac{p_{6}}{2} & \frac{p_{2} + p_{3}}{2} & \frac{p_{4}}{2} & 0
    \\
    p_{6} & p_{6} & 0 & \frac{p_{5}}{2} & \frac{p_{4}}{2} & \frac{p_{1} + p_{2}}{2} & 0
    \\
    p_{7} & p_{7} & p_{7} & 0 & 0 & 0 & \frac{1}{2} (p_{1} p_{2} + p_{1} p_{3} + p_{2} p_{3} 
    - p_{4}^{2} - p_{5}^{2} - p_{6}^{2})
  \end{pmatrix}.
\label{eq:p:matrix}
\end{equation}
Because we are interested in the case where all three triplets are non-zero, we can take unit norms $p_{1} = p_{2} = p_{3} = 1$ in the $P$-matrix \eqref{eq:p:matrix} for the determination of the orbit space boundary: then $p_{4} = \vartheta_{1}$, $p_{5} = \vartheta_{2}$, $p_{6} = \vartheta_{3}$ and $p_{7} = \vartheta_{4}$.

The orbit space is built up from semialgebraic sets, defined by equations and inequalities, of different dimensions. The interior of the orbit space corresponds to field configurations for which the gauge group is fully broken. The boundary of the orbit space corresponds to field configurations that preserve some subgroups of the full gauge group. Like a polyhedron, an orbit space typically has vertices, edges, faces and so on (which, unlike for a polyhedron, can be curved). The lower-dimensional features of the boundary correspond to higher symmetries. That is, from the orbit space vertices (0-faces) to edges (1-faces), faces (2-faces), cells (3-faces), \ldots, $k$-faces, \ldots, symmetry is broken more and more. The $k$-faces of the boundary are given by the equations
\begin{equation}
  \det P = 0, \qquad \rank P = k.
\label{eq:orbit:space:boundary}
\end{equation}
That is, the $k$-dimensional and lower minors of the $P$-matrix are non-zero while the higher ones vanish. The lower-dimensional $k$-faces correspond to field configurations with higher residual symmetry. 

In our case, the orbit space is trivial if at least one of the triplets is zero. Since, as seen from the $P$-matrix \eqref{eq:p:matrix}, taking $\rank P \leq 3$ sets at least one of the norms $p_{1}$, $p_{2}$ and $p_{3}$ to zero, the first non-trivial features of the orbit space are given by the dimension $k = 4$ principal minors taken to be zero.

%%%%%%%%%%%%%%%%%%%%%%%%%%%%%%%%%%%%%%%%%%%%%%%%%%%%%%%%%%
\begin{figure}[tbp]
\begin{center}
  \includegraphics{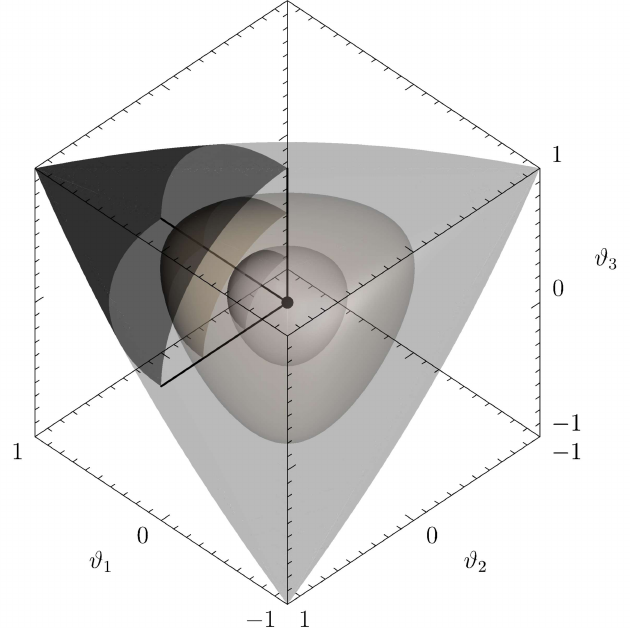}
\caption{Three-dimensional sections of the orbit space for fixed values of the orbit parameter $\vartheta_{4}$: $\vartheta_{4} = \pm 1$ gives the origin, $\vartheta_{4} = 0$ the largest bounding surface. The intersections of the orbit space with the non-negative orthant are shaded darker.}
\label{fig:orbit:space}
\end{center}
\end{figure}
%%%%%%%%%%%%%%%%%%%%%%%%%%%%%%%%%%%%%%%%%%%%%%%%%%%%%%%%%%

The boundary of the four-dimensional orbit space consists of four vertices and a cell ($3$-face).
The vertices are given by
\begin{equation}
\label{eq:vertices}
  \vartheta_{1} = \pm 1, \qquad \vartheta_{2} = \pm 1, \qquad \vartheta_{3} = \pm 1, \qquad \vartheta_{4} = 0.
\end{equation}
The cell or 3-face is given by
\begin{equation}
\label{eq:cell}
  \vartheta_{4}^{2} = 1 - \vartheta_{1}^{2} - \vartheta_{2}^{2} - \vartheta_{3}^{2} + 2 \vartheta_{1} \vartheta_{2} \vartheta_{3}.
\end{equation}
In the geometric interpretation with $\vartheta_{i}$ as angles between the triplet vectors, the right-hand-side of eq.~\eqref{eq:cell} gives the square of the volume of the parallelepiped generated by the respective unit vectors. Notice that the syzygy \eqref{eq:syzygy} reduces to the cell \eqref{eq:cell} for real fields. In this sense, the $P$-matrix, for this model, only tells us that the boundary of the orbit space is given by real fields.

The interior of the orbit space is given by $\det P > 0$, i.e. by $\vartheta_{4}^{2} < 1 - \vartheta_{1}^{2} - \vartheta_{2}^{2} - \vartheta_{3}^{2} + 2 \vartheta_{1} \vartheta_{2} \vartheta_{3}$.

The envelope of the family of three-dimensional surfaces parameterised by $\vartheta_{4}$ is bounded by the surface \eqref{eq:cell} with $\vartheta_{4} = 0$, i.e. by 
\begin{equation}
  0 = 1 - \vartheta_{1}^{2} - \vartheta_{2}^{2} - \vartheta_{3}^{2} + 2 \vartheta_{1} \vartheta_{2} \vartheta_{3}.
\label{eq:elliptope}
\end{equation} 
In convex geometry, this region is called the elliptope \cite{LAURENT1995439}.

The orbit space is pictured in figure~\ref{fig:orbit:space} for some fixed values of the orbit parameter $\vartheta_{4}$: in particular, $\vartheta_{4} = \pm 1$ gives the origin, $\vartheta_{4} = 0$ the bounding elliptope \eqref{eq:elliptope}.

%%%%%%%%%%%%%%%%%%%%%%%%%%%%%%%%%%%%%%%%%%%%%%%%%%%%%%%%%%
\subsection{Is the orbit space convex?}
\label{orbit:space:convex}

We would like to know whether the orbit space is convex. Because the potential depends linearly on $\vartheta_{i}^{2}$ and $\vartheta_{4}$, the potential minima lie on the convex hull of the orbit space: no minimum would lie in a concave part \cite{Kim:1981xu,Degee:2012sk,Heikinheimo:2017nth}. We show now that the orbit space is a convex set, i.e. the convex combinations of any two of its points lie within it.

With respect to the $\vartheta_{4}$-axis, the orbit space is composed of two symmetric halves whose boundary is given by
\begin{equation}
  \vartheta_{4\pm}(\vartheta_{i}) = \pm \sqrt{1 - \vartheta_{1}^{2} - \vartheta_{2}^{2} - \vartheta_{3}^{2} + 2 \vartheta_{1} \vartheta_{2} \vartheta_{3}}.
  \label{eq:theta:4:sol}
\end{equation}
The real functions $\vartheta_{4\pm}(\vartheta_{i})$ are defined on the convex domain of $\vartheta_{i}$ bounded by the three-dimensional elliptope \eqref{eq:elliptope}. 

\begin{figure}[tb]
\begin{center}
  \includegraphics{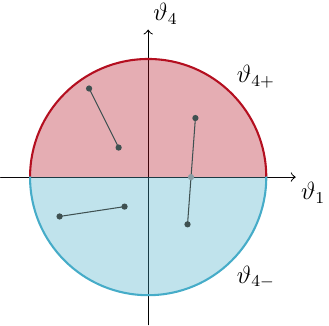}
\caption{Section of the orbit space on the $\vartheta_{1}\vartheta_{4}$-plane. The orbit space consists of two symmetric halves bounded by the graphs of the functions $\vartheta_{4+}$ and $\vartheta_{4-}$ defined on a common convex domain. Example convex combinations of orbit space points are shown in dark grey.}
\label{fig:convexity}
\end{center}
\end{figure}

A function is convex if the set of points on or above the graph of the function is a convex set. Similarly, a function is concave (convex upwards) if the set of points on or below the graph of the function is a convex set. We will show that both halves of the orbit space are convex sets and also that the convex combination of one point in the lower and another in the upper half lies in the orbit space, so the orbit space as a whole is convex. These possible combinations are illustrated in figure~\ref{fig:convexity}.

%%%%%%%%%%%%%%%%%%%%%%%%%%%%%%%%%%%%%%%%%%%%%%%%%%%%%%%%%%
%%%%%%%%%%%%%%%%%%%%%%%%%%%%%%%%%%%%%%%%%%%%%%%%%%%%%%%%%%

A twice differentiable function of several variables is convex on a convex set if and only if its Hessian matrix of second partial derivatives is positive semidefinite on the interior of the convex set. Similarly, a function is concave if its Hessian matrix is negative semidefinite.

The Hessian of $\vartheta_{4\pm}$ in eq.~\eqref{eq:theta:4:sol} as a function of $\vartheta_{i}$ is given by
\begin{equation}
\begin{split}
  H_{\pm} &= \mp (1 - \vartheta_{1}^{2} - \vartheta_{2}^{2} - \vartheta_{3}^{2} + 2 \vartheta_{1} \vartheta_{2} \vartheta_{3})^{\frac{3}{2}}
  \\
  &\times
    \begin{pmatrix}
    (\vartheta_{2}^{2} - 1) (\vartheta_{3}^{2} - 1) 
    & 
    (\vartheta_{3}^{2} - 1) (\vartheta_{3} - \vartheta_{1} \vartheta_{2})
    &
    (\vartheta_{2}^{2} - 1) (\vartheta_{2} - \vartheta_{1} \vartheta_{3})
    \\
    (\vartheta_{3}^{2} - 1) (\vartheta_{3} - \vartheta_{1} \vartheta_{2})
    &
    (\vartheta_{1}^{2} - 1) (\vartheta_{3}^{2} - 1) 
    &
    (\vartheta_{1}^{2} - 1) (\vartheta_{1} - \vartheta_{2} \vartheta_{3})
    \\
    (\vartheta_{2}^{2} - 1) (\vartheta_{2} - \vartheta_{1} \vartheta_{3})
    &
    (\vartheta_{1}^{2} - 1) (\vartheta_{1} - \vartheta_{2} \vartheta_{3})
    &
    (\vartheta_{1}^{2} - 1) (\vartheta_{2}^{2} - 1) 
  \end{pmatrix}.
\end{split}
  \label{eq:theta:4:hessian}
\end{equation}

It is easiest to study the Hessian by considering its invariants. For a $3 \times 3$ matrix $M$, its three invariants, in terms of its eigenvalues $\lambda_{i}$, $i = 1, 2, 3$, are given by $\tr M = \lambda_{1} + \lambda_{2} + \lambda_{3}$, $\tr \adj M = \lambda_{1} \lambda_{2} + \lambda_{1} \lambda_{3} + \lambda_{2} \lambda_{3}$ and $\det M = \lambda_{1} \lambda_{2} \lambda_{3}$. A positive semidefinite $M$ has $\lambda_{i} \geq 0$, so $\tr M \geq 0$, $\tr \adj M \geq 0$, $\det M \geq 0$. 
We can easily show that the Hessian $H_{-}$ is positive semidefinite on the set bounded by the elliptope \eqref{eq:elliptope}, so $\vartheta_{4-}$ is a convex function. Therefore, $H_{+} = -H_{-}$ is negative semidefinite on the set bounded by the elliptope \eqref{eq:elliptope}, so $\vartheta_{4+}$ is a concave function, i.e. it is convex upwards. That means that the convex combination of any two points on or above the $\vartheta_{4-}$ graph lies above the  graph. Similarly the convex combination of any two points on or below the $\vartheta_{4+}$ graph lies below the graph. Now let one point lie within the $\vartheta_{4-}$ graph and another within the $\vartheta_{4+}$ graph. Since the domain of both $\vartheta_{4\pm}$ functions is the same convex set, the convex combination of the $\vartheta_{i}$ coordinates of the points lies within the domain. By continuity, we can find a convex combination of $\vartheta_{i}$ for which $\vartheta_{4} = 0$: then the convex combinations with $\vartheta_{4} < 0$ lie in the lower half and the combinations with $\vartheta_{4} > 0$ in the upper half by the convexity of the halves. Therefore the orbit space as a whole is a convex set.

%%%%%%%%%%%%%%%%%%%%%%%%%%%%%%%%%%%%%%%%%%%%%%%%%%%%%%%%%%
\subsection{Potential extrema in orbit space}
\label{extrema:orbit:space}

Because the potential depends linearly on $\vartheta_{i}^{2}$, we restrict the $SU(3)_{L}$ orbit space to the non-negative, $(+,+,+)$, octant of the $\vartheta_{i}$ space, leaving $\vartheta_{4}$ unrestricted.\footnote{For the 331 potential with no $\mathbb{Z}_{2}$ symmetry, we would have to consider the whole orbit space and minima could lie not only at the boundary, but also in the interior of the orbit space, because the potential would contain terms proportional to $\vartheta_{1} \vartheta_{2}$, not just $\vartheta_{1}^{2}$ and $\vartheta_{2}^{2}$, for example. Because the interior corresponds to complex field configurations, these minima would spontaneously violate the CP symmetry.} That is, we consider the intersection of the orbit space with $\mathbb{R}_{+}^{3} \times \mathbb{R}$. Because both the $SU(3)_{L}$ orbit space and the non-negative orthant are convex regions, their intersection is also a convex region. Then the potential minima lie on the intersection of the cell of the orbit space with the first octant. In addition to the $\vartheta_{i} = 1$ vertex \eqref{eq:vertices} and the cell \eqref{eq:cell} with $0\leq \vartheta \leq 1$, we have to consider the vertices and edges of this intersection coming from the non-negative octant. Note that all these features still lie on the cell \eqref{eq:cell}, i.e. on the orbit space boundary. Its vertices are given by
\begin{equation}
\label{eq:vertex:horn:min}
  \vartheta_{1} = 0, \qquad \vartheta_{2} = 0, \qquad \vartheta_{3} = 0, \qquad \vartheta_{4} = \pm 1.
\end{equation}
These two vertices are connected by the three edges given by
\begin{align}
\label{eq:edges}
  \vartheta_{1} &= \sqrt{1 - \vartheta_{4}^{2}}, & \vartheta_{2} &= 0, & \vartheta_{3} &= 0, & -1 &\leq \vartheta_{4} \leq 1,
  \\
  \vartheta_{1} &= 0, & \vartheta_{2} &= \sqrt{1 - \vartheta_{4}^{2}}, & \vartheta_{3} &= 0, & -1 &\leq \vartheta_{4} \leq 1,
  \\
  \vartheta_{1} &= 0, & \vartheta_{2} &= 0, & \vartheta_{3} &=  \sqrt{1 - \vartheta_{4}^{2}}, & -1 &\leq \vartheta_{4} \leq 1.
\end{align}

Now that we have determined the orbit space, we can find extrema of the scalar potential \eqref{eq:V:orbit:real}. All twenty possible extrema of the scalar potential are given in table~\ref{tab:field:configurations}, listing a potential value (also serving as a name for the extremum), orbit space configuration, and a representative field configuration. They fall into eight types based on how many field norms and how many orbit variables differ from zero. Of course, one must solve the stationary point equations to obtain actual values of the orbit variables and field norms in each extremum. For given values of couplings, some extrema may not exist if the solutions for the orbit variables fell outside their physical range or field norms became imaginary.

All extrema can be presented in terms of positive real VEVs: degenerate configurations on the gauge orbit are given by $\eta \to U \eta$, $\rho \to U \rho$, $\chi \to U \chi$ with $U$ an $SU(3)_{L}$ transformation matrix. Because any vacua lie on the boundary of the orbit space which is generated by real field configurations, the immediate consequence is that there is no spontaneous CP-violation in the model at hand.

Our neutral vacuum is given by $V_{\rm EW}^{+}$ at the vertex \eqref{eq:vertex:horn:min} with $\vartheta_{4} = 1$. The value of the potential in the other neutral vertex, $V_{\rm EW}^{-}$ with $\vartheta_{4} = -1$, is always higher than in our vacuum and can never be a minimum.

The only other neutral vacua are given by $V_{O}$, $V_{\chi}$, $V_{\rho}$, $V_{\eta}$, $V_{\rho\chi}^{\perp}$, $V_{\eta\chi}^{\perp}$, $V_{\eta\chi}^{\parallel}$, $V_{\eta\rho}^{\perp}$ and $V_{\rm edge}^{1}$. The vacuum $V_{\rho\chi}^{\perp}$ is essentially the inert triplet model \cite{Dong:2013ioa}. Note that in the last neutral vacuum, $V_{\rm edge}^{1}$, the same triplet contains two neutral VEVs \cite{Diaz:2003dk}. We have $\vartheta_{2} = \vartheta_{3} = 0$ for the most general neutral VEV configuration: it is thus a basis-independent criterion to have a neutral vacuum. All the rest of the extrema break the electromagnetic $U(1)_{\rm EM}$. 

Phenomelogically viable minima must contain at least three non-zero neutral VEVs to give tree-level masses to all the particles. Therefore there are only two phenomenologically viable minima, $V_{\rm EW}^{+}$ and $V_{\rm edge}^{1}$. 

We now describe the symmetry-breaking solutions for the first six extrema with non-zero VEVs in table~\ref{tab:field:configurations}. The electroweak vacuum $V_{\rm EW}^{+}$ will be treated separately in section~\ref{sec:EW:vacuum}. As with these extrema, the minimum potential for the other cases becomes biquadratic in field norms once the solutions for the orbit variables (and Lagrange multipliers, when necessary) have been substituted. This means that in each case there is a single solution for the field norms and orbit variables. These solutions, however, are too complicated to be shown in detail.

%%%%%%%%%%%%%%%%%%%%%%%%%%%%%%%%%%%%%%%%%%%%%%%%%%%%%%%%%%
\subsubsection{Extrema along a single field}  \label{sec:onefield}

In the case of only one field acquiring a VEV, the minimum solutions and corresponding potential values are given by
\begin{align}
	\abs{\chi}^2 = -\frac{\mu_{\chi}^{2}}{2\lambda_\chi} > 0 & \implies V_{\chi} = - \frac{\mu_{\chi}^{4}}{4\lambda_\chi},\\
	\abs{\rho}^2 = -\frac{\mu_{\rho}^{2}}{2\lambda_\rho} > 0 & \implies V_{\rho} = - \frac{\mu_{\rho}^{4}}{4\lambda_\rho},\\
	\abs{\eta}^2 = -\frac{\mu_{\eta}^{2}}{2\lambda_\eta} > 0 & \implies V_{\eta} = - \frac{\mu_{\eta}^{4}}{4\lambda_\eta}.
\end{align}
Since the self-couplings of the fields must all be positive from bounded-from-below conditions, any of these vacua can only be realised with a negative $\mu_{\eta}^{2}$, $\mu_{\rho}^{2}$ or $\mu_{\eta}^{2}$. 

%%%%%%%%%%%%%%%%%%%%%%%%%%%%%%%%%%%%%%%%%%%%%%%%%%%%%%%%%%
\subsubsection{Extrema along two fields} \label{sec:twofield}

If the VEV $\abs{\eta} = 0$, then 
\begin{equation}
\abs{\rho}^2 = \frac{2 \lambda_\chi \mu_\rho^2 - (\lambda_{\rho \chi} + \lambda'_{\rho \chi} \vartheta_2^2) \mu_\chi^2}{(\lambda_{\rho \chi} + \lambda'_{\rho \chi} \vartheta_2^2)^2 - 4 \lambda_\rho \lambda_\chi} > 0, \quad
 \abs{\chi}^2 = \frac{2 \lambda_\rho \mu_\chi^2 - (\lambda_{\rho \chi} + \lambda'_{\rho \chi} \vartheta_2^2) \mu_\rho^2}{(\lambda_{\rho \chi} + \lambda'_{\rho \chi} \vartheta_2^2)^2 - 4 \lambda_\rho \lambda_\chi} > 0,
\end{equation}
with $\vartheta_2^2 = 0$ if $\lambda'_{\rho \chi}>0$ and $\vartheta_2^2 = 1$ otherwise. The potential is then

\begin{equation}
	V_{\rho\chi}^{\perp, \parallel} = \frac{\lambda_\chi \mu_\rho^4 + \lambda_\rho \mu_\chi^4 - (\lambda_{\rho \chi} + \lambda'_{\rho \chi} \vartheta_2^2 )\mu_\rho^2 \mu_\chi^2}{(\lambda_{\rho \chi} + \lambda'_{\rho \chi} \vartheta_2^2  )^2 - 4\lambda_\rho \lambda_\chi}.
\end{equation}
Similarly, with $\abs{\rho} = 0$:
\begin{align}
\abs{\eta}^2 &= \frac{2 \lambda_\chi \mu_\eta^2 - [\lambda_{\eta \chi} +  (\lambda'_{\eta \chi}-\abs{\lambda''_{\eta \chi}}) \vartheta_1^2] \mu_\chi^2}{[\lambda_{\eta \chi} +  (\lambda'_{\eta \chi}-\abs{\lambda''_{\eta \chi}}) \vartheta_1^2]^2 - 4 \lambda_\eta \lambda_\chi} > 0, \\
 \abs{\chi}^2 &= \frac{2 \lambda_\eta \mu_\chi^2 - [\lambda_{\eta \chi} +  (\lambda'_{\eta \chi}-\abs{\lambda''_{\eta \chi}}) \vartheta_1^2] \mu_\eta^2}{[\lambda_{\eta \chi} +  (\lambda'_{\eta \chi}-\abs{\lambda''_{\eta \chi}}) \vartheta_1^2]^2 - 4 \lambda_\eta \lambda_\chi} > 0,
\end{align}
with $\vartheta_1^2 = 0$ if $\lambda'_{\eta \chi} - \abs{\lambda''_{\eta \chi}}>0$ and $\vartheta_1^2 = 1$ otherwise. The potential is 
\begin{equation}
	V_{\eta\chi}^{\perp, \parallel} = \frac{\lambda_\chi \mu_\eta^4 + \lambda_\eta \mu_\chi^4 - [\lambda_{\eta \chi} +  (\lambda'_{\eta \chi}-\abs{\lambda''_{\eta \chi}}) \vartheta_1^2]\mu_\eta^2 \mu_\chi^2}{[\lambda_{\eta \chi} +  (\lambda'_{\eta \chi}-\abs{\lambda''_{\eta \chi}}) \vartheta_1^2]^2 - 4\lambda_\eta \lambda_\chi}.
\end{equation}
With $\abs{\chi} = 0$, we have
\begin{equation}
\abs{\eta}^2 = \frac{2 \lambda_\rho \mu_\eta^2 - (\lambda_{\eta \rho} +  \lambda'_{\eta \rho} \vartheta_3^2 ) \mu_\rho^2}{(\lambda_{\eta \rho} +  \lambda'_{\eta \rho} \vartheta_3^2 )^2 - 4 \lambda_\eta \lambda_\rho} > 0, \quad
 \abs{\rho}^2 = \frac{2 \lambda_\eta \mu_\rho^2 - (\lambda_{\eta \rho} +  \lambda'_{\eta \rho} \vartheta_3^2 ) \mu_\eta^2}{(\lambda_{\eta \rho} +  \lambda'_{\eta \rho} \vartheta_3^2 )^2 - 4 \lambda_\eta \lambda_\rho} > 0,
\end{equation}

with $\vartheta_3^2 = 0$ if $\lambda'_{\eta \rho}>0$ and $\vartheta_3^2 = 1$ otherwise. The potential is

\begin{equation}
	V_{\eta\rho}^{\perp, \parallel} = \frac{\lambda_\rho \mu_\eta^4 + \lambda_\eta \mu_\rho^4 - (\lambda_{\eta \rho} +  \lambda'_{\eta \rho} \vartheta_3^2 )\mu_\eta^2 \mu_\rho^2}{(\lambda_{\eta \rho} +  \lambda'_{\eta \rho} \vartheta_3^2  )^2 - 4\lambda_\eta \lambda_\rho}.
\end{equation}

%%%%%%%%%%%%%%%%%%%%%%%%%%%%%%%%%%%%%%%%%%%%%%%%%%%%%%%%%%
\begin{table}[htp]
\caption{Extrema of the potential. We list the name (as potential value), orbit configuration and a representative field configuration for each extremum. Our electroweak vacuum is $V_{\rm EW}^{+}.$}
\begin{center}
{\footnotesize
\rowcolors{1}{}{lightgray}
\begin{tabular}{llrrr}
  \hline
    Name & Orbit configuration & $\chi^{T}$ & $\rho^{T}$ & $\eta^{T}$
    \\
    \hline
    $V_{O}$ 
    &
    $\abs{\chi} = \abs{\eta} = \abs{\rho} = 0$
    &
    $(0, 0, 0)$
    &
    $(0, 0, 0)$
    &
    $(0, 0, 0)$
    \\
    $V_{\chi}$
    &
    $\abs{\eta} = \abs{\rho} = 0$
    &
    $\abs{\chi} \, (0, 0, 1)$
    &
    $(0, 0, 0)$
    &
    $(0, 0, 0)$
    \\
    $V_{\rho}$ 
    & 
    $\abs{\eta} = \abs{\chi} = 0$
    &
    $(0, 0, 0)$
    &
    $\abs{\rho} \, (0, 1, 0)$
    &
    $(0, 0, 0)$
    \\
    $V_{\eta}$
    &
    $\abs{\rho} = \abs{\chi} = 0$
    &
    $(0, 0, 0)$
    &
    $(0, 0, 0)$
    &
    $\abs{\eta} \, (1, 0, 0)$
    \\
    $V_{\rho\chi}^{\perp}$
    &
    $\abs{\eta} = 0$, $\vartheta_{2} = 0$
    &
    $\abs{\chi} \, (0, 0, 1)$
    &
    $\abs{\rho}  \, (0, 1, 0)$
    &
    $(0, 0, 0)$ 
    \\
    $V_{\rho\chi}^{\parallel}$
    &
    $\abs{\eta} = 0$, $\vartheta_{2} = 1$
    &
    $\abs{\chi} \, (0, 0, 1)$
    &
    $\abs{\rho}  \, (0, 0, 1)$
    &
    $(0, 0, 0)$ 
    \\
    $V_{\eta\chi}^{\perp}$
    &
    $\abs{\rho} = 0$, $\vartheta_{1} = 0$
    &
    $\abs{\chi} \, (0, 0, 1)$
    &
    $(0, 0, 0)$
    &
    $\abs{\eta}  \, (1, 0, 0)$
    \\
    $V_{\eta\chi}^{\parallel}$
    &
    $\abs{\rho} = 0$, $\vartheta_{1} = 1$
    &
    $\abs{\chi} \, (0, 0, 1)$
    &
    $(0, 0, 0)$
    &
    $\abs{\eta}  \, (0, 0, 1)$
    \\
    $V_{\eta\rho}^{\perp}$
    &
    $\abs{\chi} = 0$, $\vartheta_{3} = 0$
    &
    $(0, 0, 0)$
    &
    $\abs{\rho} \, (0, 1, 0)$
    &
    $\abs{\eta} \, (1, 0, 0)$
    \\
    $V_{\eta\rho}^{\parallel}$
    &
    $\abs{\chi} = 0$, $\vartheta_{3} = 1$
    &
    $(0, 0, 0)$
    &
    $\abs{\rho} \, (1, 0, 0)$
    &
    $\abs{\eta} \, (1, 0, 0)$
    \\
    $V_{\rm tip}$
    &
    $\vartheta_{i}^{2} = 1$, $\vartheta_{4} = 0$
    &
    $\abs{\chi} \, (0, 0, 1)$
    &
    $\abs{\rho} \, (0, 0, 1)$
    &
    $\abs{\eta}  \, (0, 0, 1)$
    \\
    $V_{\rm EW}^{+}$
    &
    $\vartheta_{i} = 0$, $\vartheta_{4} = 1$
    &
    $\frac{1}{\sqrt{2}}v_{\chi} (0, 0, 1)$
    &
    $\frac{1}{\sqrt{2}}v_{\rho} (0, 1, 0)$
    &
    $\frac{1}{\sqrt{2}}v_{\eta} (1, 0, 0)$
    \\
    $V_{\rm EW}^{-} $
    &
    $\vartheta_{i} = 0$, $\vartheta_{4} = -1$
    &
    $\abs{\chi} \, (1, 0, 0)$
    &
    $\abs{\rho} \, (0, 1, 0)$
    &
    $\abs{\eta} \, (0, 0, 1)$
    \\
    $V_{\rm edge}^{1}$
    &
    $\begin{aligned}
      \vartheta_{1} &= \sqrt{1 - \vartheta_{4}^{2}}, \\
      \vartheta_{2} &= \vartheta_{3} = 0, \\
      -1 &< \vartheta_{4} < 1
    \end{aligned}$
    &
    $\abs{\chi} \, (\sqrt{1 - \vartheta_{4}^{2}}, 0, \vartheta_{4})$
    &
    $\abs{\rho} \, (0, 1, 0)$
    &
    $\abs{\eta} \, (1, 0, 0)$
    \\
    $V_{\rm edge}^{2}$
    &
    $\begin{aligned}
      \vartheta_{2} &= \sqrt{1 - \vartheta_{4}^{2}}, \\
      \vartheta_{1} &= \vartheta_{3} = 0, \\
      -1 &< \vartheta_{4} < 1
    \end{aligned}$
    &
    $\abs{\chi} \, (0, 1, 0)$
    &
    $\abs{\rho} \, (\vartheta_{4}, \sqrt{1 - \vartheta_{4}^{2}}, 0)$
    &
    $\abs{\eta} \, (0, 0, 1)$
    \\
    $V_{\rm edge}^{3}$
    &
    $\begin{aligned}
      \vartheta_{3} &= \sqrt{1 - \vartheta_{4}^{2}}, \\
      \vartheta_{1} &= \vartheta_{2} = 0, \\
      -1 &< \vartheta_{4} < 1
    \end{aligned}$
    &
    $\abs{\chi} \,  (0, 0, 1)$
    &
    $\abs{\rho} \, (0, 1, 0)$
    &
    $\abs{\eta} \, (\vartheta_{4}, \sqrt{1 - \vartheta_{4}^{2}}, 0)$
    \\
    $V_{\rm face}^{12}$
    &
    $\begin{aligned}
      \vartheta_{1} &= \sqrt{1 - \vartheta_{2}^{2} - \vartheta_{4}^{2}}, \\
      \vartheta_{3} &= 0, 0 < \vartheta_{2} < 1, \\
      -1 &< \vartheta_{4} < 1
    \end{aligned}$
    &
    $\abs{\chi} \, (\sqrt{1- \vartheta_{1}^{2} - \vartheta_{4}^{2}}, 
    \vartheta_{4}, \vartheta_{1})$
    &
    $\abs{\rho} \, (1, 0, 0)$
    &
    $\abs{\eta} \, (0, 0, 1)$
    \\
    $V_{\rm face}^{13}$
    &
    $\begin{aligned}
      \vartheta_{3} &= \sqrt{1 - \vartheta_{1}^{2} - \vartheta_{4}^{2}}, \\
      \vartheta_{2} &= 0, 0 < \vartheta_{1} < 1, \\
      -1 &< \vartheta_{4} < 1
    \end{aligned}$
    &
    $\abs{\chi} \, (1, 0, 0)$
    &
    $\abs{\rho} \, (0, 0, 1)$
    &
    $\abs{\eta} \, (\sqrt{1- \vartheta_{3}^{2} - \vartheta_{4}^{2}}, \vartheta_{4}, \vartheta_{3})$
    \\

    $V_{\rm face}^{23}$
    &
    $\begin{aligned}
      \vartheta_{3} &= \sqrt{1 - \vartheta_{2}^{2} - \vartheta_{4}^{2}}, \\
      \vartheta_{1} &= 0, 0 < \vartheta_{2} < 1, \\
      -1 &< \vartheta_{4} < 1
    \end{aligned}$
    &
    $\abs{\chi} \, (0, 0, 1)$
    &
    $\abs{\rho} \, (\sqrt{1- \vartheta_{2}^{2} - \vartheta_{4}^{2}}, \vartheta_{4}, \vartheta_{2})$
    &
    $\abs{\eta} \, (1, 0, 0)$
    \\
    $V_{\rm cell}$
    &
    $\begin{aligned}
      &1 - \vartheta_{1}^{2} - \vartheta_{2}^{2} - \vartheta_{3}^{2} \\
      &+ 2 \vartheta_{1} \vartheta_{2} \vartheta_{3} - \vartheta_{4}^{2} = 0, \\ 
      &0 < \vartheta_{i} < 1
    \end{aligned}$
    &
    $(\chi_{1}, 0, \chi_{3})$
    &
    $(\rho_{1}, \rho_{2}, 0)$
    &
    $(0, \eta_{2}, \eta_{3})$
\end{tabular}
}
\end{center}
\label{tab:field:configurations}
\end{table}%
%%%%%%%%%%%%%%%%%%%%%%%%%%%%%%%%%%%%%%%%%%%%%%%%%%%%%%%%%%

%%%%%%%%%%%%%%%%%%%%%%%%%%%%%%%%%%%%%%%%%%%%%%%%%%%%%%%%%%
\section{Constraints}
\label{sec:exp:constr}

%%%%%%%%%%%%%%%%%%%%%%%%%%%%%%%%%%%%%%%%%%%%%%%%%%%%%%%%%%
\subsection{Perturbative unitarity}
\label{sec:exp:unit}

Perturbative unitarity arises from the unitarity of the scattering matrix for scalar two-to-two scattering amplitudes. Considering only the zeroth partial wave, the $S$-matrix is given by
\begin{equation}
  a_{0}^{ba} = \sqrt{\frac{4 \abs{\mathbf{p}_{b}} \abs{\mathbf{p}_{a}}}{2^{\delta_{a}} 2^{\delta_{b}} s}}
  \int_{-1}^{1} d(\cos \theta) \mathcal{M}_{ba} (\cos \theta),
\end{equation}
with a pair of scalars $a$ scattering to the pair $b$ with the matrix element $\mathcal{M}_{ba} (\cos \theta)$. The angle $\theta$ is between the incoming three-momenta $\mathbf{p}_{a}$ and the outgoing $\mathbf{p}_{b}$ in the centre-of-mass frame and the Mandelstam variable $s = (p_{1} + p_{2})^{2}$. The Kronecker $\delta_{a}$ is unity if the particles in pair $a$ are identical and zero otherwise (likewise for $\delta_{b}$ and $b$). The eigenvalues $a_{0}^{i}$ of the scattering matrix must satisfy
\begin{equation}
  \abs{\Re a_{0}^{i}} \leq \frac{1}{2}.
\end{equation}
In the limit of high energy with $s \to \infty$, only quartic couplings contribute to scattering. We compute the scattering matrix for all possible two-to-two processes $S_{1}S_{2} \to S_{3}S_{4}$ of scalar bosons, including Goldstone bosons. In this limit, the $a_{0}^{ba}$ matrix element, with $a \equiv S_{1} S_{2}$ and $b \equiv S_{3} S_{4}$, is given by
\begin{equation}
  a_{0}^{ba} = \frac{1}{16 \pi} \frac{1}{\sqrt{2^{\delta_{S_{1} S_{2}}} 2^{\delta_{S_{3} S_{4}}}}} \frac{\partial^{4} V}{\partial S_{1} \partial S_{2} \partial S_{3}^{*} \partial S_{4}^{*}}.
\end{equation}
Considering all scatterings of zero-charge, single-charge and double-charge initial and final states, we obtain the perturbative unitarity constraints
\begin{equation}
\begin{aligned}
  \abs{\lambda_{\eta}} &< \pi, 
  &
  \abs{\lambda_{\rho}} &< \pi,  
  & 
  \abs{\lambda_{\chi}} &< \pi, 
  \\
  \abs{\lambda_{\eta\rho}} &< 8 \pi, 
  &
  \abs{\lambda_{\eta\rho} \pm \lambda'_{\eta\rho}} &< 8 \pi,
  &
  \abs{\lambda_{\eta\rho} + 3 \lambda'_{\eta\rho}} &< 8 \pi,
  \\
  \abs{\lambda_{\rho\chi}} &< 8 \pi,
  &
  \abs{\lambda_{\eta\chi} \pm \lambda'_{\eta\chi}} &< 8 \pi, 
  & 
  \abs{\lambda_{\eta\chi} \pm \lambda''_{\eta\chi}} &< 8 \pi, 
  \\
  \abs{\lambda_{\rho\chi} \pm \lambda'_{\rho\chi}} &< 8 \pi,
  &
  \!\!\!\!
  \abs{\lambda_{\rho\chi} + 3 \lambda'_{\rho\chi}} &< 8 \pi,
  &
  \!\!\!\!
  \abs{\lambda_{\eta\chi} + 3 \lambda'_{\eta\chi} \pm 4 \lambda''_{\eta\chi}} &< 8 \pi,
  \\ 
   \span \span \span \span
  \abs{\lambda_{\eta} + \lambda_{\chi} \pm \sqrt{\lambda_{\eta\chi}^{\prime\prime 2} + (\lambda_{\eta} - \lambda_{\chi})^{2}}} &< 8 \pi,
\end{aligned}
\end{equation}
and the solutions of the cubic equations
\begin{equation}
\begin{split}
  0 &= x^{3} 
  - 8 (\lambda_{\eta} + \lambda_{\rho} + \lambda_{\chi}) x^{2}
  + [64 (\lambda_{\eta} \lambda_{\rho} + \lambda_{\eta} \lambda_{\chi} + \lambda_{\rho} \lambda_{\chi}) 
  \\
  &- (3 \lambda_{\eta\rho} + \lambda'_{\eta\rho})^{2}
  - (3 \lambda_{\eta\chi} + \lambda'_{\eta\chi})^{2} - (3 \lambda_{\rho\chi} + \lambda'_{\rho\chi})^{2}] x - 512 \lambda_{\eta} \lambda_{\rho} \lambda_{\chi}  
  \\
  & + 8 \lambda_{\eta} (3 \lambda_{\rho\chi} + \lambda'_{\rho\chi})^{2} + 8 \lambda_{\rho} (3 \lambda_{\eta\chi} + \lambda'_{\eta\chi})^{2} + 8 \lambda_{\chi} (3 \lambda_{\eta\rho} + \lambda'_{\eta\rho})^{2}
  \\
  &- 2 (3 \lambda_{\eta\rho} + \lambda'_{\eta\rho}) (3 \lambda_{\eta\chi} + \lambda'_{\eta\chi}) (3 \lambda_{\rho\chi} + \lambda'_{\rho\chi}),
  \\
  0&= y^{3} - 2 (\lambda_{\eta} + \lambda_{\rho} + \lambda_{\chi}) y^{2} 
  + (4 \lambda_{\eta} \lambda_{\rho} + 4 \lambda_{\eta} \lambda_{\chi} + 4 \lambda_{\rho} \lambda_{\chi} - \lambda_{\eta\rho}^{\prime 2}  - \lambda_{\eta\chi}^{\prime 2} - \lambda_{\rho\chi}^{\prime 2}) y
  \\
  &+ 2 (\lambda_{\rho} \lambda_{\eta\chi}^{\prime 2} + \lambda_{\eta} \lambda_{\rho\chi}^{\prime 2} + \lambda_{\chi} \lambda_{\eta\rho}^{\prime 2} 
- 4 \lambda_{\eta} \lambda_{\rho} \lambda_{\chi} - \lambda'_{\eta\rho} \lambda'_{\eta\chi} \lambda'_{\rho\chi}),
\end{split}
\end{equation}
satisfy $x_{i} < 8 \pi$, $y_{i} < 8 \pi$, $i = 1,2,3$. Previous perturbative unitarity constraints obtained in ref.~\cite{Costantini:2020xrn} consider a generic parameter $\beta$ in which case the scattering matrices are smaller.

%%%%%%%%%%%%%%%%%%%%%%%%%%%%%%%%%%%%%%%%%%%%%%%%%%%%%%%%%%
\subsection{Boundedness of the potential from below}
\label{sec:bfb}

In order for the scalar potential to make physical sense, it must be bounded from below, i.e. the minimum of the potential energy must be finite. In the limit of large field values, we can disregard the dimensionful mass terms and the trilinear term, and impose conditions solely on the scalar quartic couplings. The orbit variable $\vartheta_{4}$ associated with the trilinear term does not enter the quartic potential and the quartic potential depends monotonously on $\vartheta_{i}^{2}$. Therefore the quartic potential $V_{4}$ has to be minimised at extremal values of $\vartheta_{i}^{2}$ which lie on the intersection of the non-negative orthant with the three-dimensional $\vartheta_{i}$ surface \eqref{eq:elliptope}. More specifically, they must be minimised at the convex hull of this intersection. For that reason, it is not necessary to separately minimise the quartic potential on the $\vartheta_{i}$ axes nor on the coordinate planes, because they are already accounted for.

Since the quartic potential is biquadratic, copositivity  can be used to derive the boundedness-from-below constraints \cite{Kannike:2012pe}. We complete the necessary conditions given in \cite{Sanchez-Vega:2018qje} with the copositivity condition on the cell \eqref{eq:cell}, presenting full necessary and sufficient conditions for the potential to be bounded from below. For the coupling matrix
\begin{equation}
  \Lambda = 
  \begin{pmatrix}
    \lambda_{\eta} &
    \frac{1}{2} (\lambda_{\eta\rho} + \lambda'_{\eta\rho} \vartheta_{3}^{2}) &
    \frac{1}{2} [\lambda_{\eta\chi} + (\lambda'_{\eta\chi} - \abs{\lambda''_{\eta\chi}}) \vartheta_{1}^{2}]
    \\
    \frac{1}{2} (\lambda_{\eta\rho} + \lambda'_{\eta\rho} \vartheta_{3}^{2}) &
    \lambda_{\rho} &
    \frac{1}{2} (\lambda_{\rho\chi} + \lambda'_{\rho\chi} \vartheta_{2}^{2})
    \\
    \frac{1}{2} [\lambda_{\eta\chi} + (\lambda'_{\eta\chi} - \abs{\lambda''_{\eta\chi}}) \vartheta_{1}^{2}] &
    \frac{1}{2} (\lambda_{\rho\chi} + \lambda'_{\rho\chi} \vartheta_{2}^{2}) &
    \lambda_{\chi}
  \end{pmatrix},
\end{equation}
the copositivity constraints
\begin{align}
  \lambda_{\eta} &> 0, & \lambda_{\rho} &> 0, & \lambda_{\chi} &> 0, 
  \label{eq:1:field:BfB}
  \\
  \bar{\lambda}_{\eta\rho} &\equiv \frac{1}{2} (\lambda_{\eta\rho} + \lambda'_{\eta\rho} \vartheta_{3}^{2}) + \sqrt{\lambda_{\eta} \lambda_{\rho}} > 0,
  \span \span \span \span
  \label{eq:2:field:BfB:eta:rho}
  \\
  \bar{\lambda}_{\eta\chi} &\equiv \frac{1}{2} (\lambda_{\eta\chi} + (\lambda'_{\eta\chi} - \abs{\lambda''_{\eta\chi}}) \vartheta_{1}^{2}) + \sqrt{\lambda_{\eta} \lambda_{\chi}} > 0,
  \span \span \span \span
    \label{eq:2:field:BfB:eta:chi}
  \\
  \bar{\lambda}_{\rho\chi} &\equiv \frac{1}{2} (\lambda_{\rho\chi} + \lambda'_{\rho\chi} \vartheta_{2}^{2}) + \sqrt{\lambda_{\rho} \lambda_{\chi}} > 0,
  \span \span \span \span
  \label{eq:2:field:BfB:rho:chi}
  \\
  \sqrt{\lambda_{\eta} \lambda_{\rho} \lambda_{\chi}} + (\lambda_{\rho\chi} + \lambda'_{\rho\chi} \vartheta_{2}^{2}) \sqrt{\lambda_{\eta}} + [\lambda_{\eta\chi} + (\lambda'_{\eta\chi} - \abs{\lambda''_{\eta\chi}}) \vartheta_{1}^{2}] \sqrt{\lambda_{\rho}}  \span \span \span \span \span
  \notag
  \\
  \span \span \span \span
  + (\lambda_{\eta\rho} + \lambda'_{\eta\rho} \vartheta_{3}^{2}) \sqrt{\lambda_{\chi}} + \sqrt{2 \bar{\lambda}_{\eta\rho} \bar{\lambda}_{\eta\chi} \bar{\lambda}_{\rho\chi}} &> 0,
\label{eq:3:field:BfB}
\end{align}
must hold at the origin $\vartheta_{i} = 0$, at the vertices given by the unit vectors $\vartheta_{i} = 1$, $\vartheta_{j \neq i} = 0$, $i, j = 1,2,3$ and by $\vartheta_{1} = \vartheta_{2} = \vartheta_{3} = 1$, and at the elliptope surface \eqref{eq:elliptope} for positive $\vartheta_{i}$.

At the surface \eqref{eq:elliptope}, it is too cumbersome to minimise the three-field condition \eqref{eq:3:field:BfB} --- where the field norms have been eliminated --- with respect to $\vartheta_{i}$. In this case, it is easier to minimise the potential explicitly on the sphere $\abs{\eta}^{2} + \abs{\rho}^{2} + \abs{\chi}^{2} = 1$ instead \cite{Kannike:2016fmd}. We enforce the constraint \eqref{eq:elliptope} and the sphere condition with two Lagrange multipliers. Minimising the potential, we obtain solutions for the orbit variables $\vartheta_{i}$, field norms, and the Lagrange multipliers. Requiring that the solutions stay within their physical ranges, we obtain the copositivity constraints on the cell, given by 
\begin{equation}
  0 < \vartheta_{1}^{2} < 1 \land 0 < \vartheta_{2}^{2} < 1 \land 0 < \vartheta_{3}^{2} < 1 
  \land \abs{\eta}^{2} > 0 \land \abs{\rho}^{2} > 0 \land \abs{\chi}^{2} >0 \implies V_{4} > 0,
\label{eq:3:field:BfB:elliptope}
\end{equation}
where strict inequalities are used because for equalities the conditions are reduced to previous ones.

The full necessary and sufficient conditions for the potential to be bounded from below are given by eqs. \eqref{eq:1:field:BfB}, \eqref{eq:2:field:BfB:eta:rho}, \eqref{eq:2:field:BfB:eta:chi}, \eqref{eq:2:field:BfB:rho:chi}, \eqref{eq:3:field:BfB} at the origin and vertices, and eq. \eqref{eq:3:field:BfB:elliptope}.

%%%%%%%%%%%%%%%%%%%%%%%%%%%%%%%%%%%%%%%%%%%%%%%%%%%%%%%%%%
\subsection{Metastability of the electroweak vacuum}
\label{sec:EW:vacuum:metastability}

For absolute vacuum stability, we must require that the neutral electroweak symmetry breaking vacuum be the global extremum of the scalar potential. If it occurs that the electroweak vacuum is not global, then we could tunnel from our vacuum into the global one. For a not too fast tunnelling rate, our vacuum could be metastable. We use the FindBounce code \cite{Guada:2018jek,Guada:2020xnz} to compute the Euclidean action to determine the tunnelling rate.

The bubble nucleation rate per unit time and volume is approximately given by 
\begin{equation}
  \Gamma \approx R_{0}^{4} e^{-S_{E}},
\end{equation}
where $R_{0}$ is the radius of the critical bubble and $S_{E}$ is the Euclidean action. The vacuum is metastable if no bubble has nucleated within the past lightcone with volume $V$ and lifetime $T$ of the Universe:
\begin{equation}
  \Gamma V T \approx 0.15 \, H_{0}^{-4} \, \Gamma < 1,
\end{equation}
where $H_{0} = 67.4~\mathrm{(km/s)/Mpc} = 1.44 \times 10^{-42}~\mathrm{GeV}$ is the Hubble constant.

In our numerical studies, the only minima that endanger absolute stability are the $V_{\rho}$, $V_{\eta}$ and $V_{\eta\rho}$. For this reason it suffices, at least in first approximation, to consider tunnelling only in the field space of real $\chi_{3}$, $\rho_{2}$, $\eta_{1}$.

%%%%%%%%%%%%%%%%%%%%%%%%%%%%%%%%%%%%%%%%%%%%%%%%%%%%%%%%%%
\section{Electroweak vacuum}
\label{sec:EW:vacuum}

We assume that in our neutral electroweak minimum the fields have the minimal VEV structure necessary to give masses to all the particles, given by
\begin{equation}
\langle \rho\rangle =
\frac{1}{\sqrt{2}}\begin{pmatrix}
0\\
v_\rho\\
0
\end{pmatrix},
\quad
\langle \eta\rangle =
\frac{1}{\sqrt{2}}\begin{pmatrix}
v_\eta\\
0\\
0
\end{pmatrix},\quad
\langle \chi\rangle =
\frac{1}{\sqrt{2}}\begin{pmatrix}
0\\
0\\
v_\chi
\end{pmatrix}
\end{equation}
with $v_{\eta}^{2} + v_{\rho}^{2} = v^{2} = (246.22~\mathrm{GeV})^{2}$ to ensure a correct $SU(3)_L\times U(1)_X\to U(1)_{\rm EM}$ symmetry breaking. Because the $\chi$ triplet is responsible for the first step of symmetry breaking, we have $v_{\chi} \gg v_{\eta}, v_{\rho}$. Note that the tree-level fermion masses require two $SU(2)_L$-breaking VEVs in two different triplets to avoid degeneracies in the quark mass matrices.

 The VEV $v_\chi > 3.6$~TeV  due to the LEP bound on the electroweak precision $\rho$ parameter \cite{ParticleDataGroup:2024cfk}. Different hierarchies between $v_\rho, v_\eta,v_\chi$ and $f$ have been studied in \cite{Pinheiro:2022bcs}.

Solving the minimisation equations in the neutral vacuum for the mass terms, we obtain 
\begin{align}
\label{eq:mass:ew:min}
  \mu_{\eta}^{2} &= \frac{f}{2} \frac{v_{\rho} v_{\chi}}{v_{\eta}}
  - \lambda_{\eta} v_{\eta}^{2} - \frac{1}{2} \lambda_{\eta\rho} v_{\rho}^{2} 
  - \frac{1}{2} \lambda_{\eta\chi} v_{\chi}^{2},
  \\\label{eq:mass:ew:min2}
  \mu_{\rho}^{2} &= \frac{f}{2} \frac{v_{\eta} v_{\chi}}{v_{\rho}}
  - \lambda_{\rho} v_{\rho}^{2} - \frac{1}{2}  \lambda_{\eta\rho} v_{\eta}^{2} 
  - \frac{1}{2}  \lambda_{\rho\chi} v_{\chi}^{2},
  \\\label{eq:mass:ew:min3}
  \mu_{\chi}^{2} &= \frac{f}{2} \frac{v_{\eta} v_{\rho}}{v_{\chi}}
  - \lambda_{\chi} v_{\chi}^{2} - \frac{1}{2}  \lambda_{\eta\chi} v_{\eta}^{2} 
  - \frac{1}{2} \lambda_{\rho\chi} v_{\rho}^{2}.
\end{align}

To study mass eigenstates, we consider the neutral components of the scalar triplets~\eqref{eq:triplets} in terms of real and imaginary parts:
\begin{equation}
\begin{aligned}
  \rho_2^0 &= \frac{1}{\sqrt{2}}(h_1+i\xi_1), 
  & 
  \eta_1^0 &= \frac{1}{\sqrt{2}}(h_2+i\xi_2),
  &
  \chi_3^0 &= \frac{1}{\sqrt{2}}(h_3+i\xi_3),
  \notag
  \\
  \chi_1^0 &=\frac{1}{\sqrt{2}}(h_4+i\xi_4),
  &
  \eta_3^0 &= \frac{1}{\sqrt{2}}(h_5+i\xi_5).
\end{aligned}
\end{equation}

\subsection{CP-even scalars}
Since the fields $\eta_{1}^{0}$, $\rho_{2}^{0}$ and $\chi_{3}^{0}$ get VEVs, their real components $h_1, h_2$ and $h_3$ all mix with each other. Their mass matrix in the basis $(h_1,h_2,h_3)$ is given by
\begin{equation}
\label{eq:M:h}
M_{H}^2=\begin{pmatrix}
\frac{f v_\eta v_\chi}{2 v_\rho} + 2 \lambda_\rho v_\rho^2 &
-\frac{f v_\chi}{2}+ \lambda_{\eta\rho} v_\eta v_\rho &
-\frac{f v_\eta}{2}+ \lambda_{\rho\chi} v_\rho v_\chi \\
-\frac{f v_\chi}{2}+ \lambda_{\eta\rho} v_\eta v_\rho &
\frac{f v_\rho v_\chi}{2 v_\eta} + 2 \lambda_\eta v_\eta^2 &
-\frac{f v_\rho}{2}+ \lambda_{\eta\chi} v_\eta v_\chi \\
-\frac{f v_\eta}{2}+ \lambda_{\rho\chi} v_\rho v_\chi  &
-\frac{f v_\rho}{2}+ \lambda_{\eta\chi} v_\eta v_\chi  &
\frac{f v_\eta v_\rho}{2 v_\chi} + 2 \lambda_\chi v_\chi^2 
\end{pmatrix}.
\end{equation}

Because the fields $\eta_{3}^{0}$ and $\chi_{1}^{0}$ do not get VEVs, their real components $h_4$ and $h_5$ mix separately, with the mass matrix in the basis $(h_4,h_5)$ given by
\begin{equation}
M_{H'}^2=\begin{pmatrix}
\frac{v_\eta [f v_\rho +(\lambda'_{\eta\chi}-\abs{\lambda''_{\eta\chi}}) v_\eta v_\chi]}{2v_\chi} &
\frac{1}{2} [f v_\rho + (\lambda'_{\eta\chi}-\abs{\lambda''_{\eta\chi}}) v_\eta v_\chi]\\
\frac{1}{2} [f v_\rho + (\lambda'_{\eta\chi}-\abs{\lambda''_{\eta\chi}}) v_\eta v_\chi] &
\frac{v_\chi [f v_\rho + (\lambda'_{\eta\chi}-\abs{\lambda''_{\eta\chi}}) v_\eta v_\chi]}{2v_\eta}  
\end{pmatrix}.
\end{equation}
The two mass matrices are diagonalised as
\begin{equation}
  U_{H}^T M_{H}^2 U_{H}=\diag(m_h^2, m_{H_1}^2,m_{H_2}^2) 
  \qquad \text{and} \qquad 
  U_{H'}^T M_{H'}^2 U_{H'}=\diag(m_{H_3}^2, 0),
\end{equation}
where $m_h = 125.11$~GeV is the mass of the SM-like Higgs and $m_{H_1}$, $m_{H_2}$, $m_{H_3}$ are heavy. 

The diagonalisation matrix $U_{H}$ can be parametrised in terms of mixing angles $\alpha_{12}$, $\alpha_{23}$ and $\alpha_{13}$; we abbreviate $\sin \alpha_{ij} \equiv s_{ij}$ and $\cos \alpha_{ij} \equiv c_{ij}$:
\begin{equation}
U_{H}=
\begin{pmatrix}
1 & 0 & 0\\
0 & c_{23} & s_{23}\\
0 & -s_{23} & c_{23}
\end{pmatrix}
\begin{pmatrix}
c_{13} & 0 & s_{13}\\
0 & 1 & 0\\
-s_{13} & 0 & c_{13}
\end{pmatrix}
\begin{pmatrix}
c_{12} & s_{12} & 0\\
-s_{12} & c_{12} & 0\\
0 & 0 & 1
\end{pmatrix}.
\end{equation}
Note that we use a different convention for the mixing matrix $U_{H}$, similar to the CKM and PMNS matrices, than ref.~\cite{Costantini:2020xrn}.

Assuming $v_\chi, f\gg v_\rho,v_\eta$, the mass matrix (\ref{eq:M:h}) can be block-diagonalised. This yields  
\begin{equation} \label{eq:vevaprox}
\cos \alpha_{12} \approx \frac{v_\rho}{\sqrt{v_\rho^2+v_\eta^2}},\quad \textrm{and}\quad
\sin \alpha_{12} \approx -\frac{v_\eta}{\sqrt{v_\rho^2+v_\eta^2}}
\end{equation}
at leading order. The masses of the heavy fields, to leading order, are
\begin{equation}
m_{H_1}^2 \approx \frac{f(v_\eta^2+v_\rho^2)v_\chi}{2 v_\eta v_\rho},
\qquad
\text{and}
\qquad
m_{H_2}^2 \approx 2 \lambda_\chi v_\chi^2 .
\end{equation}

The assumption $v_\chi, f\gg v_\rho,v_\eta$ corresponds to the decoupling limit where the couplings of $125$ GeV
Higgs become SM-like. The Yukawa couplings of $h$ become those of the SM.\footnote{There is always a presence of flavour-violating couplings in the quark sector in non-sequential 331-models, but these tend to zero as $v_\chi\to\infty$.}
 
%%%%%%%%%%%%%%%%%%%%%%%%%%%%%%%%%%%%%%%%%%%%%%%%%%%%%%%%%%

\subsection{CP-odd scalars}

Similarly to scalars, we have two different sets of pseudoscalars that do not mix with each other.
The fields $\xi_1, \xi_2$ and $\xi_3$ mix: their mass matrix in the basis $(\xi_1,\xi_2,\xi_3)$ is
\begin{equation}
M_{A}^2=\begin{pmatrix}
\frac{f v_\eta v_\chi}{2 v_\rho} &
\frac{f v_\chi}{2} &
\frac{f v_\eta}{2}\\
\frac{f v_\chi}{2} &
\frac{f v_\rho v_\chi}{2 v_\eta} &
\frac{f v_\rho}{2}\\
\frac{f v_\eta}{2} &
\frac{f v_\rho}{2} &
\frac{f v_\eta v_\rho}{2 v_\chi} 
\end{pmatrix}.
\end{equation}
The fields $\xi_4$ and $\xi_5$ mix separately, with the mass matrix in the basis $(\xi_4,\xi_5)$ given by
\begin{equation}
M_{A'}^2=\begin{pmatrix}
\frac{v_\eta[f v_\rho + (\lambda'_{\eta\chi}+ \abs{\lambda''_{\eta\chi}}) v_\eta v_\chi] }{2v_\chi} &
-\frac{1}{2}[f v_\rho + (\lambda'_{\eta\chi}+ \abs{\lambda''_{\eta\chi}}) v_\eta v_\chi]\\
-\frac{1}{2}[f v_\rho + (\lambda'_{\eta\chi}+ \abs{\lambda''_{\eta\chi}}) v_\eta v_\chi] &
\frac{v_\chi [f v_\rho + (\lambda'_{\eta\chi}+ \abs{\lambda''_{\eta\chi}}) v_\eta v_\chi]}{2v_\eta}  
\end{pmatrix}.
\end{equation}
These mass matrices are diagonalised as
\begin{equation}
U_{A}^T M_{A}^2 U_{A}=\diag(m_{A_1}^2, 0,0) \qquad \text{and} \qquad
U_{A'}^T M_{A'}^2 U_{A'}=\diag(m_{A_2}^2, 0).
\end{equation}

%%%%%%%%%%%%%%%%%%%%%%%%%%%%%%%%%%%%%%%%%%%%%%%%%%%%%%%%%%

\subsection{Charged scalars}
For the charged scalars, as well, there are two sets of fields that do not mix with each other. The field $\rho_1^+$ mixes with $\eta_2^+$, and $\rho_3^+$ mixes with $\chi_2^+$. The mass matrices in the bases $(\rho_1^+,\eta_2^+)$ and $(\rho_3^+,\chi_2^+)$ are 
\begin{equation}
M_{C}^2=
\begin{pmatrix}
\frac{v_\eta(f v_\chi+\lambda'_{\eta\rho}v_\eta v_\rho)}{2v_\rho} &
\frac{1}{2}(f v_\chi +\lambda'_{\eta\rho}v_\eta v_\rho)\\
\frac{1}{2}(f v_\chi +\lambda'_{\eta\rho}v_\eta v_\rho) &
\frac{v_\rho(f v_\chi+\lambda'_{\eta\rho}v_\eta v_\rho)}{2v_\eta}
\end{pmatrix},
\end{equation}
and
\begin{equation}
M_{C'}^2=
\begin{pmatrix}
\frac{v_\chi(f v_\eta+\lambda'_{\rho\chi}v_\rho v_\chi)}{2v_\rho} &
\frac{1}{2}(f v_\eta +\lambda'_{\rho\chi}v_\rho v_\chi)\\
\frac{1}{2}(f v_\eta +\lambda'_{\rho\chi}v_\rho v_\chi) &
\frac{v_\rho(f v_\eta+\lambda'_{\rho\chi}v_\rho v_\chi)}{2v_\chi}
\end{pmatrix},
\end{equation}
which can be diagonalised as
\begin{equation}
U_{C}^T M_{C}^2 U_{C}=\diag(m_{H_1^+}^2, 0)
\qquad
\text{and}
\qquad
U_{C'}^T M_{C'}^2 U_{C'}=\diag(m_{H_2^+}^2, 0).
\end{equation}

%%%%%%%%%%%%%%%%%%%%%%%%%%%%%%%%%%%%%%%%%%%%%%%
\subsection{Parametrisation}\label{sec:parametrisation}

Now we can exchange the original potential parameters for physical parameters: VEVs, masses and mixing angles. Note that the unprimed scalar couplings also depend on the mixing angles,
\begin{equation}
\begin{split}
\lambda_\rho =& 
\frac{
c_{\alpha_{12}}^2 c_{\alpha_{13}}^2 m_h^2 
+c_{\alpha_{13}}^2(1-c_{\alpha_{12}}^2)m_{H_1}^2
+(1-c_{\alpha_{13}}^2)m_{H_2}^2
-\frac{f v_\eta v_\chi}{2v_\rho}}{2v_\rho^2},
\\
%%%%%%%%%%%%%%%%%%%%%%%%%%
\lambda_\eta =&
\frac{(s_{\alpha_{12}}c_{\alpha_{23}}+c_{\alpha_{12}}s_{\alpha_{13}}s_{\alpha_{23}})^2 m_h^2}{2v_\eta^2}
+\frac{(c_{\alpha_{12}}c_{\alpha_{23}}-s_{\alpha_{12}}s_{\alpha_{13}}s_{\alpha_{23}})^2 m_{H_1}^2}{2v_\eta^2}
\\
&+\frac{c_{\alpha_{13}}^2 s_{\alpha_{23}}^2  m_{H_2}^2}{2v_\eta^2}
-\frac{f v_\rho v_\chi}{4 v_\eta^3},
\\
%%%%%%%%%%%%%%%%%%%%%%%%%%
\lambda_\chi =& 
\frac{(c_{\alpha_{12}}s_{\alpha_{13}}c_{\alpha_{23}}
-s_{\alpha_{12}}s_{\alpha_{23}})^2 m_h^2}{2v_\chi^2}
+\frac{(s_{\alpha_{12}}s_{\alpha_{13}}c_{\alpha_{23}}+c_{\alpha_{12}}s_{\alpha_{23}})^2 m_{H_1}^2}{2v_\chi^2}
\\
&+\frac{c_{\alpha_{13}}^2 c_{\alpha_{23}}^2  m_{H_2}^2}{2v_\chi^2}
-\frac{f v_\eta v_\rho}{4 v_\chi^3},
\\
%%%%%%%%%%%%%%%%%%%%%%%%%%
\lambda_{\eta\rho} =&
\frac{(s_{2\alpha_{12}}c_{\alpha_{13}}c_{\alpha_{23}}
+c_{\alpha_{12}}^2 s_{2\alpha_{13}}s_{\alpha_{23}})
(m_{H_1}^2-m_h^2)
+s_{2\alpha_{13}}s_{\alpha_{23}}(m_{H_2}^2-m_{H_1}^2)+f v_\chi}{2v_\eta v_\rho},\\
%%%%%%%%%%%%%%%%%%%%%%%%%%
\lambda_{\rho\chi} =&\frac{
(c_{\alpha_{12}}^2 s_{2\alpha_{13}} c_{\alpha_{23}}-s_{2\alpha_{12}} c_{\alpha_{13}}s_{\alpha_{23}})(m_{H_1}^2-m_h^2)+s_{2\alpha_{13}}c_{\alpha_{23}}(m_{H_2}^2-m_{H_1}^2)
+f v_\eta}{2v_\rho v_\chi},
\\
%%%%%%%%%%%%%%%%%%%%%%%%%%
\lambda_{\eta\chi} =&
\frac{1}{8v_\eta v_\chi} [2c_{\alpha_{13}}^2s_{2\alpha_{23}}(2m_{H_2}^2-m_h^2-m_{H_1}^2)+(c_{2\alpha_{12}}s_{2\alpha_{23}}(c_{2\alpha_{13}}-3)
\\
&-4c_{2\alpha_{23}}s_{2\alpha_{12}}s_{\alpha_{13}})(m_{H_1}^2-m_h^2)+4fv_\rho],
\end{split}
\end{equation}
whereas the primed couplings only depend on the masses and VEVs:
\begin{equation}
\begin{split}
%%%%%%%%%%%%%%%%%%%%%%%%%%
\lambda'_{\eta\rho} =&
2 \frac{ m_{H^+_1}^2}{v_\eta^2+v_\rho^2}
-\frac{2m_{A_1}^2 v_\chi^2}{v_\rho^2 v_\chi^2 +v_\eta^2 v_\rho^2+ v_\eta^2 v_\chi^2},
\\
\lambda'_{\rho\chi} =&
\frac{2 m_{H^+_2}^2}{v_\rho^2+v_\chi^2}
-\frac{2m_{A_1}^2 v_\eta^2}{v_\rho^2 v_\chi^2 +v_\eta^2 v_\rho^2+ v_\eta^2 v_\chi^2},
\\
\lambda'_{\eta\chi} =&
\frac{m_{A_2}^2+m_{H_3}^2}{v_\eta^2+v_\chi^2}
-\frac{2m_{A_1}^2 v_\rho^2}{v_\rho^2 v_\chi^2 +v_\eta^2 v_\rho^2+ v_\eta^2 v_\chi^2},
\\
\lambda''_{\eta\chi} =&\frac{m_{H_3}^2-m_{A_2}^2}{v_\eta^2+v_\chi^2},
\\
 f =& \frac{2 m_{A_1}^2 v_\eta v_\rho v_\chi}{v_\rho^2 v_\chi^2 +v_\eta^2 v_\rho^2+ v_\eta^2 v_\chi^2}.
\end{split}
\end{equation}

%%%%%%%%%%%%%%%%%%%%%%%%%%%%%%%%%%%%%%%%%%%%%%%%%%%%%%%%%%
\section{Stability of the electroweak vacuum}
\label{sec:stab}

Of the twenty possible extrema in table~\ref{tab:field:configurations}, one is our vacuum $V_{\rm EW}^{+}$ or $V_{\rm EW}$ for short. Its counterpart $V_{\rm EW}^{-}$ with $\vartheta_{4} = -1$ can never be a minimum. For the remaining eighteen extrema, we must compare their potential energy with $V_{\rm EW}$ to determine whether our vacuum is global and absolutely stable. Although we have analytical conditions for globality, it is difficult to determine simple conditions for the parameter space. For that reason, we resort to a combination of numerical and analytical calculations.

We first perform a Markov Chain Monte Carlo (MCMC) scan to determine which extrema can be deeper than our vacuum in some part of the parameter space. We varied VEV $v_\chi$, mixing angles, and all unknown masses, giving a total of 11 independent parameters. The VEV and masses were chosen in the 1--100~TeV range and the angle $\sin \alpha_{12}$ was limited to negative values to ensure positive VEVs. We used the relations in eq. \eqref{eq:vevaprox} and $v_{\eta}^{2} + v_{\rho}^{2} = v^{2} = (246.22~\mathrm{GeV})^{2}$ to determine the VEVs of the other two fields. In the first part of the scan, we searched for viable extrema for each vacuum case separately. We preferred points that were closer to satisfying the conditions for the extremum to be physical, such as having real norms and orbital parameters in the allowed range. After finding a handful of viable points, the second step of the scan focused on searching for points where that vacuum becomes the global minimum, starting from the points found in the previous step. For this, we favoured points with the smallest relative difference from the actual global minimum, continuing until the desired vacuum overtook the status of global minimum. If no such point was found, the search was abandoned after 20 000 iterations. The second step was then repeated, totaling about 200 000 iterations before the search was fully abandoned.
 
  As a result, we find only $V_{\eta}$, $V_{\rho}$ and $V_{\eta\rho}$ as possibly deeper than our vacuum. We proceed to analyse stability of our vacuum in the leading order of $f \sim v_\chi \gg v_\eta, v_\rho$ and establish clear conditions regarding these extrema.

The simplest case is at the origin of field space, where  the value of the potential is $V_{O} = 0$. For the origin to be a minimum, the masses of the particles given by $\mu_{\eta}^{2}$, $\mu_{\rho}^{2}$ and $\mu_{\chi}^{2}$ must be positive. This could only be achieved with $f \gg v_\chi$, which is not compatible with the electroweak vacuum as a minimum because it would make $\det M_{H}^{2} < 0$. For $f \sim v_{\chi}$ we always have $\mu_{\eta}^{2} < 0$, $\mu_{\rho}^{2} < 0$ or $\mu_{\chi}^{2} < 0$.

The situation simplifies for the case when we can take $\vartheta_{i} = 0$. If $\lambda'_{\eta \chi} - \abs{\lambda''_{\eta \chi}}>0$, $\lambda'_{\rho \chi}>0$,  and $\lambda'_{\eta \rho}>0$ then all terms with $\vartheta_{i}$ orbit variables give positive contributions to the potential. Therefore in this case all extrema with nonzero $\vartheta_{i}$ can be discarded, as our vacuum is always deeper. Then we only need to consider the six extrema discussed in sections~\ref{sec:onefield} and \ref{sec:twofield}. We now derive simple approximations for these extrema to be deeper than our vacuum.

First of all, the $V_\chi$ extremum, which almost coinsides with our $V_{\rm EW}$, is in fact always shallower than ours: 
 \begin{equation}
 	V_{\rm EW} - V_\chi = - \frac{m_h^2 v_h^2}{8} + \mathcal{O}\left( \frac{v_h^6}{v^2_\chi} \right).
 \end{equation}
There are two other vacua along single field directions, for which
 \begin{align}
 	V_{\rm EW} - V_\rho &= \frac{v_\chi^4}{16} \left[ \frac{\left(f \frac{v_\eta}{v_\rho v_\chi} - \lambda_{\rho \chi}\right)^2}{\lambda_\rho} - 4 \lambda_\chi \right]
 	 + \mathcal{O}\left( v_h v_\chi^3 \right), \\
 	 V_{\rm EW} - V_\eta &= \frac{v_\chi^4}{16} \left[ \frac{\left(f \frac{v_\rho}{v_\eta v_\chi} - \lambda_{\eta \chi}\right)^2}{\lambda_\eta} - 4 \lambda_\chi \right]
 	 + \mathcal{O}\left( v_h v_\chi^3 \right).
 \end{align}

As described in section~\ref{sec:onefield}, $\mu_\rho^2<0$ ($\mu_\eta^2<0$) must be satisfied for $V_\rho$ ($V_\eta$) to be minimum. This is possible only with positive $\lambda_{\rho \chi}$ ($\lambda_{\eta \chi}$), leading to simple conditions for $V_{\rm EW}<V_\rho, V_\eta$:

 \begin{equation}
 	\lambda_{\rho \chi} < f \frac{v_\eta}{v_\rho v_\chi} + 2\sqrt{\lambda_\rho \lambda_\chi}, \quad
 	\lambda_{\eta \chi} < f \frac{v_\rho}{v_\eta v_\chi} + 2\sqrt{\lambda_\eta \lambda_\chi}.
 \end{equation}

Comparing the $V_{\rho\chi}$ and $V_{\eta\chi}$ extrema to the EW vacuum gives

\begin{align}
	V_{\rm EW} - V_{\rho\chi} \approx V_{\chi} - V_{\rho\chi} &= \frac {(\lambda_{\rho \chi} \mu_\chi^2 - 2 \lambda_\chi \mu_\rho^2)^2}
	{4 \lambda_\chi (4\lambda_\rho \lambda_\chi - \lambda_{\rho \chi}^2) }, \\
	V_{\rm EW} - V_{\eta\chi} \approx V_{\chi} - V_{\rho\chi} &= \frac {(\lambda_{\eta \chi} \mu_\chi^2 - 2 \lambda_\chi \mu_\eta^2)^2}
	{4 \lambda_\chi (4\lambda_\eta \lambda_\chi - \lambda_{\eta \chi}^2) }.
\end{align}
We see that $V_{\rho\chi}$ ($V_{\eta\chi}$) could be lower than EW only if $4\lambda_\rho \lambda_\chi > \lambda_{\rho \chi}^2$ ($4\lambda_\eta \lambda_\chi > \lambda_{\eta \chi}^2$), but this gives rise to unphysical $\abs{\rho}^2<0$ ($\abs{\eta}^2<0$) in this region.

Finally, for the $V_{\eta\rho}$ extremum, the conditions for $V_{\rm EW}<V_{\eta\rho}$ are given by
\begin{equation}
 \lambda_{\eta \rho} > \frac{\delta_{\eta \chi}\delta_{\rho \chi} - \sqrt{4 \lambda_\eta \lambda_\chi-\delta_{\eta \chi}^2} \sqrt{4 \lambda_\rho \lambda_\chi-\delta_{\rho \chi}^2}}{2\lambda_\chi} \, \lor \,
 \lambda_{\eta \rho} \mu_\rho^2 < 2 \lambda_\rho \mu_\eta^2 \, \lor \,
 \lambda_{\eta \rho} \mu_\eta^2 < 2 \lambda_\eta \mu_\rho^2,
\end{equation}
where $\delta_{\eta \chi} = \lambda_{\eta \chi} - f v_\rho/(v_\eta v_\chi)$ and $\delta_{\rho \chi} = \lambda_{\rho \chi} - f v_\eta/(v_\rho v_\chi)$.

For any of $\lambda'_{\eta \chi} - \abs{\lambda''_{\eta \chi}} < 0$, $\lambda'_{\rho \chi} < 0$, and $\lambda'_{\eta \rho} <0$, the EW vacuum $V_{\rm EW}$ typically remained the global minimum in the parameter region tested by the MCMC scan. In case of extrema $V_\rho$ and $V_\eta$, the solutions became unphysical, giving $\abs{\rho}^2<0$ and $\abs{\eta}^2<0$, respectively. However, $V_{\eta \rho}$ could become global.

%%%%%%%%%%%%%%%%%%%%%%%%%%%%%%%%%%%%%%%%%%%%%%%%%%%%%%%%%%
\begin{figure}[tbp]
\begin{center}
    \includegraphics{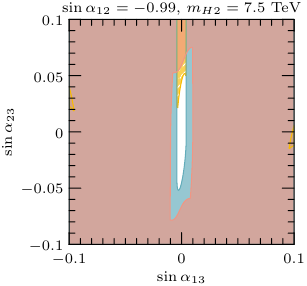}~
  \includegraphics{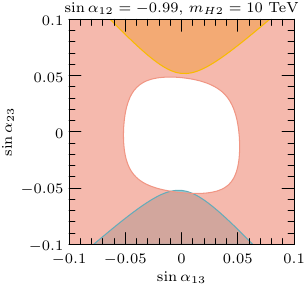}~
  \includegraphics{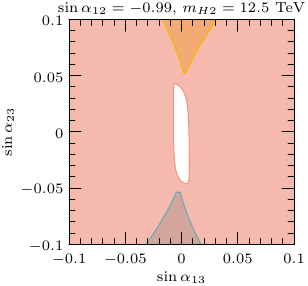}
  \\
  \includegraphics{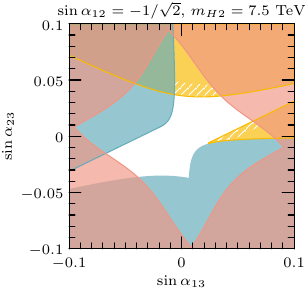}~
  \includegraphics{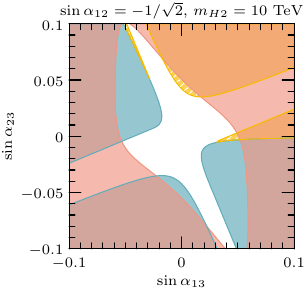}~
  \includegraphics{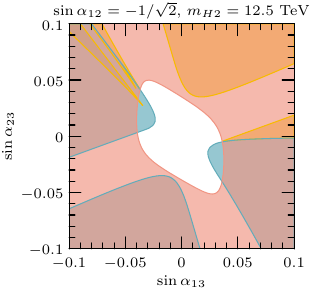}
  \\
  \includegraphics{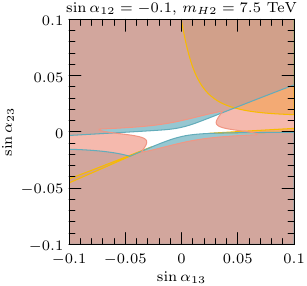}~
  \includegraphics{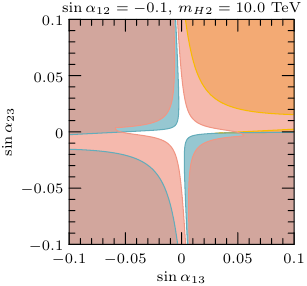}~
  \includegraphics{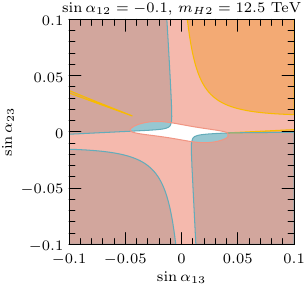}
\caption{Parameter space of the 331 model with three triplets. Regions where unitarity is violated are shown in red, where the potential is not bounded from below in blue, where the electroweak vacuum is not global in yellow (metastable in hatched yellow). Regions in white satisfy all constraints.}
\label{fig:param:space:1}
\end{center}
\end{figure}
%%%%%%%%%%%%%%%%%%%%%%%%%%%%%%%%%%%%%%%%%%%%%%%%%%%%%%%%%%

%%%%%%%%%%%%%%%%%%%%%%%%%%%%%%%%%%%%%%%%%%%%%%%%%%%%%%%%%%
\begin{figure}[tbp]
\begin{center}
  \includegraphics{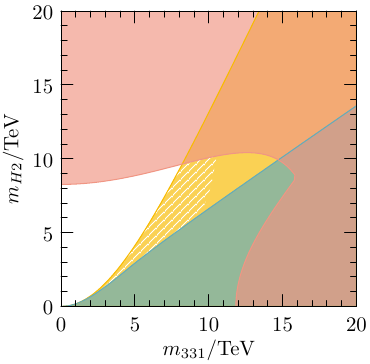}
\caption{Parameter space of the 331 model with $\sin \alpha_{12} = -1/\sqrt{2}$, $\sin \alpha_{13} = 0$, $\sin \alpha_{23} = 0.05$. Regions where unitarity is violated are shown in red, where the potential is not bounded from below in blue, where the electroweak vacuum is not global in yellow (metastable in hatched yellow). Regions in white satisfy all constraints.}
\label{fig:param:space:2}
\end{center}
\end{figure}
%%%%%%%%%%%%%%%%%%%%%%%%%%%%%%%%%%%%%%%%%%%%%%%%%%%%%%%%%%

In figures \ref{fig:param:space:1} and \ref{fig:param:space:2}, we illustrate, for $v_{\chi} = 10$ TeV, the typical parameter space. We take all the scalars, except for the SM-like Higgs boson, to be heavy with a common mass scale $m_{331}$. Note that unitarity already requires $m_{A_{1}} \approx m_{H_{1}^{+}} \approx m_{H_{1}}$. We also vary the mass of the $H_{2}$ scalar since it has a strong influence on whether the electroweak vacuum is global. For figure \ref{fig:param:space:1}, we choose $m_{331} = 10$~TeV and plot the parameter space on the $\sin \alpha_{23}$ vs. $\sin \alpha_{13}$ plane for $m_{H2} = 0.75 \, m_{331}$, $m_{H2} = 1.00 \, m_{331}$, $m_{H2} = 1.25 \, m_{331}$ and $\sin \alpha_{12} = -\sin (\pi/32) \approx -0.1$, $\sin \alpha_{12} = -1/\sqrt{2}$, $\sin \alpha_{12} = - \sin (\pi/2 - \pi/32) \approx -0.995$. For figure \ref{fig:param:space:2}, we fix $\sin \alpha_{12} = -1/\sqrt{2}$, $\sin \alpha_{13} = 0$, $\sin \alpha_{23} = 0.05$ and plot the constraints on the $m_{H2}$ vs. $m_{331}$ plane. In the figures, regions where unitarity is violated are shown in red, where potential is not bounded from below in blue, where electroweak vacuum is not global in yellow. Regions in white satisfy all constraints. The electroweak vacuum is metastable in regions in hatched yellow. The globality constraint takes analytically into account all the extrema. We calculate the metastability bound only in regions where the unitarity and boundedness-from-below constraints are satisfied. In the yellow region, the electroweak vacuum is not global because the $V_{\eta}$ and $V_{\rho}$ extrema are deeper.

%%%%%%%%%%%%%%%%%%%%%%%%%%%%%%%%%%%%%%%%%%%%%%%%%%%%%%%%%%
\section{Conclusions}
\label{sec:concl}

We find, in the three-triplet 331 model with a $\mathbb{Z}_{2}$ symmetry and a trilinear $\mathbb{Z}_{2}$-breaking soft term, all the possible extrema of the scalar potential, and study conditions for the electroweak vacuum to be global. To our knowledge, we are the first to study the full vacuum structure of this potential. This is relevant for phenomenology because in regions of parameter space that are otherwise allowed, our vacuum may turn out to be unstable. When our vacuum is not global, we also calculate the tunnelling rate, to see where it is metastable.

Finding the extrema of the scalar potential is simpler in the orbit space as it exchanges a large number of real field degrees of freedom for a smaller space of gauge invariants albeit with a non-trivial shape. We determine the shape of the orbit space by the $P$-matrix method. The  orbit space, pictured in figure~\ref{fig:orbit:space}, finds a simple geometrical interpretation. It is then straightforward to find all the potential extrema, listed in table~\ref{tab:field:configurations} together with corresponding minimal field configurations. We also find a basis-invariant condition for a vacuum to be neutral.

 Of course, the potential has to be bounded from below, for which we present the full necessary and sufficient conditions, completing the conditions given in ref.~\cite{Sanchez-Vega:2018qje}. Other main constraints are given by perturbative unitarity.

We consider in detail a typical parameter space in the limit of large $v_{\chi} \approx f$ and all heavy masses, except one, equal to a common $m_{331}$ mass scale. The bounds from unitarity, boundedness-from-below and (meta)stability are shown in figures~\ref{fig:param:space:1} and \ref{fig:param:space:2}. We find that the electroweak vacuum may not be global if the mixing of the $\eta$ and $\rho$ triplets with the triplet $\chi$ is non-zero. The stability condition can strongly constrain available parameter space as seen from figure~\ref{fig:param:space:2}. A large part of the parameter space where the electroweak vacuum is not global is, however, still metastable. On the other  hand, in the part of the typical parameter space where this mixing is negligible, the model is unitary, the potential bounded from below and the electroweak vacuum is global.

The ancillary Wolfram Mathematica notebook contains constraints from perturbative unitarity, boundedness from below, vacuum stability, and the parametrisation of the couplings.

The orbit space analysis could be extended to the gauge and fermion sector. The masses of gauge bosons and fermions can also be expressed via orbit variables so as to give everything in the same formalism. The same orbit space could also be used to find minima of the potential without the $\mathbb{Z}_{2}$ symmetry.

%%%%%%%%%%%%%%%%%%%%%%%%%%%%%%%%%%%%%%%%%%%%%%%%%%%%%%%%%%
\section*{Acknowledgements}

We thank Antonio Costantini for discussions and Vin\'{i}cius Padovani for reading the manuscript. This work was supported by the Estonian Research Council grants TEM-TA23, PRG1677, RVTT3, RVTT7, TARISTU24-TK10, TARISTU24-TK3, and the CoE TK 202 ``Foundations of the Universe''.

%%%%%%%%%%%%%%%%%%%%%%%%%%%%%%%%%%%%%%%%%%%%%%%%%%%%%%%%%%
\bibliography{331_orbit_space}

\providecommand{\href}[2]{#2}\begingroup\raggedright\begin{thebibliography}{10}

\bibitem{Georgi:1978bv}
H.~Georgi and A.~Pais, \emph{{Generalization of Gim: Horizontal and Vertical
  Flavor Mixing}}, \href{https://doi.org/10.1103/PhysRevD.19.2746}{\emph{Phys.
  Rev. D} {\bfseries 19} (1979) 2746}.

\bibitem{Singer:1980sw}
M.~Singer, J.W.F.~Valle and J.~Schechter, \emph{{Canonical Neutral Current
  Predictions From the Weak Electromagnetic Gauge Group SU(3) X $u$(1)}},
  \href{https://doi.org/10.1103/PhysRevD.22.738}{\emph{Phys. Rev. D} {\bfseries
  22} (1980) 738}.

\bibitem{Valle:1983dk}
J.W.F.~Valle and M.~Singer, \emph{{Lepton Number Violation With Quasi Dirac
  Neutrinos}}, \href{https://doi.org/10.1103/PhysRevD.28.540}{\emph{Phys. Rev.
  D} {\bfseries 28} (1983) 540}.

\bibitem{Montero:1992jk}
J.C.~Montero, F.~Pisano and V.~Pleitez, \emph{{Neutral currents and GIM
  mechanism in SU(3)-L x U(1)-N models for electroweak interactions}},
  \href{https://doi.org/10.1103/PhysRevD.47.2918}{\emph{Phys. Rev. D}
  {\bfseries 47} (1993) 2918}
  [\href{https://arxiv.org/abs/hep-ph/9212271}{{\ttfamily hep-ph/9212271}}].

\bibitem{Foot:1994ym}
R.~Foot, H.N.~Long and T.A.~Tran, \emph{{$SU(3)_L \otimes U(1)_N$ and $SU(4)_L
  \otimes U(1)_N$ gauge models with right-handed neutrinos}},
  \href{https://doi.org/10.1103/PhysRevD.50.R34}{\emph{Phys. Rev. D} {\bfseries
  50} (1994) R34} [\href{https://arxiv.org/abs/hep-ph/9402243}{{\ttfamily
  hep-ph/9402243}}].

\bibitem{Long:1995ctv}
H.N.~Long, \emph{{The 331 model with right handed neutrinos}},
  \href{https://doi.org/10.1103/PhysRevD.53.437}{\emph{Phys. Rev. D} {\bfseries
  53} (1996) 437} [\href{https://arxiv.org/abs/hep-ph/9504274}{{\ttfamily
  hep-ph/9504274}}].

\bibitem{Long:1996rfd}
H.N.~Long, \emph{{SU(3)-L x U(1)-N model for right-handed neutrino neutral
  currents}}, \href{https://doi.org/10.1103/PhysRevD.54.4691}{\emph{Phys. Rev.
  D} {\bfseries 54} (1996) 4691}
  [\href{https://arxiv.org/abs/hep-ph/9607439}{{\ttfamily hep-ph/9607439}}].

\bibitem{Pleitez:1994pu}
V.~Pleitez, \emph{{New fermions and a vector - like third generation in SU(3)
  (C) x SU(3) (L) x U(1) ($N$) models}},
  \href{https://doi.org/10.1103/PhysRevD.53.514}{\emph{Phys. Rev. D} {\bfseries
  53} (1996) 514} [\href{https://arxiv.org/abs/hep-ph/9412304}{{\ttfamily
  hep-ph/9412304}}].

\bibitem{Long:1997vbr}
H.N.~Long, \emph{{Scalar sector of the 3 3 1 model with three Higgs triplets}},
  \href{https://doi.org/10.1142/S0217732398001959}{\emph{Mod. Phys. Lett. A}
  {\bfseries 13} (1998) 1865}
  [\href{https://arxiv.org/abs/hep-ph/9711204}{{\ttfamily hep-ph/9711204}}].

\bibitem{Dong:2013ioa}
P.V.~Dong, T.P.~Nguyen and D.V.~Soa, \emph{{3-3-1 model with inert scalar
  triplet}}, \href{https://doi.org/10.1103/PhysRevD.88.095014}{\emph{Phys. Rev.
  D} {\bfseries 88} (2013) 095014}
  [\href{https://arxiv.org/abs/1308.4097}{{\ttfamily 1308.4097}}].

\bibitem{Pisano:1992bxx}
F.~Pisano and V.~Pleitez, \emph{{An SU(3) x U(1) model for electroweak
  interactions}}, \href{https://doi.org/10.1103/PhysRevD.46.410}{\emph{Phys.
  Rev. D} {\bfseries 46} (1992) 410}
  [\href{https://arxiv.org/abs/hep-ph/9206242}{{\ttfamily hep-ph/9206242}}].

\bibitem{Frampton:1992wt}
P.H.~Frampton, \emph{{Chiral dilepton model and the flavor question}},
  \href{https://doi.org/10.1103/PhysRevLett.69.2889}{\emph{Phys. Rev. Lett.}
  {\bfseries 69} (1992) 2889}.

\bibitem{Foot:1992rh}
R.~Foot, O.F.~Hernandez, F.~Pisano and V.~Pleitez, \emph{{Lepton masses in an
  SU(3)-L x U(1)-N gauge model}},
  \href{https://doi.org/10.1103/PhysRevD.47.4158}{\emph{Phys. Rev. D}
  {\bfseries 47} (1993) 4158}
  [\href{https://arxiv.org/abs/hep-ph/9207264}{{\ttfamily hep-ph/9207264}}].

\bibitem{Tonasse:1996cx}
M.D.~Tonasse, \emph{{The Scalar sector of 3-3-1 models}},
  \href{https://doi.org/10.1016/0370-2693(96)00481-9}{\emph{Phys. Lett. B}
  {\bfseries 381} (1996) 191}
  [\href{https://arxiv.org/abs/hep-ph/9605230}{{\ttfamily hep-ph/9605230}}].

\bibitem{Nguyen:1998ui}
T.A.~Nguyen, N.A.~Ky and H.N.~Long, \emph{{The Higgs sector of the minimal 3 3
  1 model revisited}},
  \href{https://doi.org/10.1142/S0217751X00000136}{\emph{Int. J. Mod. Phys. A}
  {\bfseries 15} (2000) 283}
  [\href{https://arxiv.org/abs/hep-ph/9810273}{{\ttfamily hep-ph/9810273}}].

\bibitem{Ponce:2002sg}
W.A.~Ponce, Y.~Giraldo and L.A.~Sanchez, \emph{{Minimal scalar sector of 3-3-1
  models without exotic electric charges}},
  \href{https://doi.org/10.1103/PhysRevD.67.075001}{\emph{Phys. Rev. D}
  {\bfseries 67} (2003) 075001}
  [\href{https://arxiv.org/abs/hep-ph/0210026}{{\ttfamily hep-ph/0210026}}].

\bibitem{Dong:2006mg}
P.V.~Dong, H.N.~Long, D.T.~Nhung and D.V.~Soa, \emph{{SU(3)(C) x SU(3)(L) x
  U(1)(X) model with two Higgs triplets}},
  \href{https://doi.org/10.1103/PhysRevD.73.035004}{\emph{Phys. Rev. D}
  {\bfseries 73} (2006) 035004}
  [\href{https://arxiv.org/abs/hep-ph/0601046}{{\ttfamily hep-ph/0601046}}].

\bibitem{Ferreira:2011hm}
J.G.~Ferreira, Jr, P.R.D.~Pinheiro, C.A.d.S.~Pires and P.S.R.~da~Silva,
  \emph{{The Minimal 3-3-1 model with only two Higgs triplets}},
  \href{https://doi.org/10.1103/PhysRevD.84.095019}{\emph{Phys. Rev. D}
  {\bfseries 84} (2011) 095019}
  [\href{https://arxiv.org/abs/1109.0031}{{\ttfamily 1109.0031}}].

\bibitem{Cogollo:2013mga}
D.~Cogollo, F.S.~Queiroz and P.~Vasconcelos, \emph{{Flavor Changing Neutral
  Current Processes in a Reduced Minimal Scalar Sector}},
  \href{https://doi.org/10.1142/S0217732314501739}{\emph{Mod. Phys. Lett. A}
  {\bfseries 29} (2014) 1450173}
  [\href{https://arxiv.org/abs/1312.0304}{{\ttfamily 1312.0304}}].

\bibitem{Dong:2014esa}
P.V.~Dong, N.T.K.~Ngan and D.V.~Soa, \emph{{Simple 3-3-1 model and implication
  for dark matter}},
  \href{https://doi.org/10.1103/PhysRevD.90.075019}{\emph{Phys. Rev. D}
  {\bfseries 90} (2014) 075019}
  [\href{https://arxiv.org/abs/1407.3839}{{\ttfamily 1407.3839}}].

\bibitem{DeConto:2015eia}
G.~De~Conto, A.C.B.~Machado and V.~Pleitez, \emph{{Minimal 3-3-1 model with a
  spectator sextet}},
  \href{https://doi.org/10.1103/PhysRevD.92.075031}{\emph{Phys. Rev. D}
  {\bfseries 92} (2015) 075031}
  [\href{https://arxiv.org/abs/1505.01343}{{\ttfamily 1505.01343}}].

\bibitem{Diaz:2003dk}
R.A.~Diaz, R.~Martinez and F.~Ochoa, \emph{{The Scalar sector of the SU(3)(c) x
  SU(3)(L) x U(1)(X) model}},
  \href{https://doi.org/10.1103/PhysRevD.69.095009}{\emph{Phys. Rev. D}
  {\bfseries 69} (2004) 095009}
  [\href{https://arxiv.org/abs/hep-ph/0309280}{{\ttfamily hep-ph/0309280}}].

\bibitem{Costantini:2020xrn}
A.~Costantini, M.~Ghezzi and G.M.~Pruna, \emph{{Theoretical constraints on the
  Higgs potential of the general $331$ model}},
  \href{https://doi.org/10.1016/j.physletb.2020.135638}{\emph{Phys. Lett. B}
  {\bfseries 808} (2020) 135638}
  [\href{https://arxiv.org/abs/2001.08550}{{\ttfamily 2001.08550}}].

\bibitem{Giraldo:2009yi}
Y.~Giraldo, W.A.~Ponce and L.A.~Sanchez, \emph{{Stability of the Scalar
  Potential and Symmetry Breaking in the Economical 3-3-1 Model}},
  \href{https://doi.org/10.1140/epjc/s10052-009-1101-4}{\emph{Eur. Phys. J. C}
  {\bfseries 63} (2009) 461} [\href{https://arxiv.org/abs/0907.1696}{{\ttfamily
  0907.1696}}].

\bibitem{Giraldo:2011gd}
Y.~Giraldo and W.A.~Ponce, \emph{{Scalar Potential Without Cubic Term in 3-3-1
  Models Without Exotic Electric Charges}},
  \href{https://doi.org/10.1140/epjc/s10052-011-1693-3}{\emph{Eur. Phys. J. C}
  {\bfseries 71} (2011) 1693}
  [\href{https://arxiv.org/abs/1107.3260}{{\ttfamily 1107.3260}}].

\bibitem{Dorsch:2024ddk}
G.C.~Dorsch, A.A.~Louzi, B.L.~S{\'a}nchez-Vega and A.C.D.~Viglioni,
  \emph{{Vacuum stability in the one-loop approximation of a 331 model}},
  \href{https://doi.org/10.1140/epjc/s10052-024-12840-4}{\emph{Eur. Phys. J. C}
  {\bfseries 84} (2024) 471}
  [\href{https://arxiv.org/abs/2402.00155}{{\ttfamily 2402.00155}}].

\bibitem{Maniatis:2006fs}
M.~Maniatis, A.~von Manteuffel, O.~Nachtmann and F.~Nagel, \emph{{Stability and
  symmetry breaking in the general two-Higgs-doublet model}},
  \href{https://doi.org/10.1140/epjc/s10052-006-0016-6}{\emph{Eur. Phys. J. C}
  {\bfseries 48} (2006) 805}
  [\href{https://arxiv.org/abs/hep-ph/0605184}{{\ttfamily hep-ph/0605184}}].

\bibitem{Ivanov:2006yq}
I.P.~Ivanov, \emph{{Minkowski space structure of the Higgs potential in 2HDM}},
  \href{https://doi.org/10.1103/PhysRevD.75.035001}{\emph{Phys. Rev. D}
  {\bfseries 75} (2007) 035001}
  [\href{https://arxiv.org/abs/hep-ph/0609018}{{\ttfamily hep-ph/0609018}}].

\bibitem{Ivanov:2007de}
I.P.~Ivanov, \emph{{Minkowski space structure of the Higgs potential in 2HDM.
  II. Minima, symmetries, and topology}},
  \href{https://doi.org/10.1103/PhysRevD.77.015017}{\emph{Phys. Rev. D}
  {\bfseries 77} (2008) 015017}
  [\href{https://arxiv.org/abs/0710.3490}{{\ttfamily 0710.3490}}].

\bibitem{Huitu:2024nap}
K.~Huitu, N.~Koivunen, T.~K\"arkk\"ainen and S.~Mondal, \emph{{On the family
  discrimination in 331-model}},
  \href{https://doi.org/10.1007/JHEP03(2025)109}{\emph{JHEP} {\bfseries 03}
  (2025) 109} [\href{https://arxiv.org/abs/2409.13013}{{\ttfamily
  2409.13013}}].

\bibitem{Pal:1994ba}
P.B.~Pal, \emph{{The Strong CP question in SU(3)(C) x SU(3)(L) x U(1)(N)
  models}}, \href{https://doi.org/10.1103/PhysRevD.52.1659}{\emph{Phys. Rev. D}
  {\bfseries 52} (1995) 1659}
  [\href{https://arxiv.org/abs/hep-ph/9411406}{{\ttfamily hep-ph/9411406}}].

\bibitem{Sanchez-Vega:2016dwe}
B.L.~S\'anchez-Vega, E.R.~Schmitz and J.C.~Montero, \emph{{New constraints on
  the 3-3-1 model with right-handed neutrinos}},
  \href{https://doi.org/10.1140/epjc/s10052-018-5626-2}{\emph{Eur. Phys. J. C}
  {\bfseries 78} (2018) 166}
  [\href{https://arxiv.org/abs/1612.03827}{{\ttfamily 1612.03827}}].

\bibitem{Pinheiro:2022bcs}
J.P.~Pinheiro and C.A.~de~S.~Pires, \emph{{On the Higgs spectra of the 3-3-1
  model}}, \href{https://doi.org/10.1016/j.physletb.2022.137584}{\emph{Phys.
  Lett. B} {\bfseries 836} (2023) 137584}
  [\href{https://arxiv.org/abs/2210.05426}{{\ttfamily 2210.05426}}].

\bibitem{Huitu:2017ukq}
K.~Huitu and N.~Koivunen, \emph{{Froggatt-Nielsen mechanism in a model with
  $SU(3)_c\times SU(3)_L \times U(1)_X$ gauge group}},
  \href{https://doi.org/10.1103/PhysRevD.98.011701}{\emph{Phys. Rev. D}
  {\bfseries 98} (2018) 011701}
  [\href{https://arxiv.org/abs/1706.09463}{{\ttfamily 1706.09463}}].

\bibitem{Cao:2016uur}
Q.-H.~Cao and D.-M.~Zhang, \emph{{Collider Phenomenology of the 3-3-1 Model}},
  \href{https://arxiv.org/abs/1611.09337}{{\ttfamily 1611.09337}}.

\bibitem{Alves:2022hcp}
A.~Alves, L.~Duarte, S.~Kovalenko, Y.M.~Oviedo-Torres, F.S.~Queiroz and
  Y.S.~Villamizar, \emph{{Constraining 3-3-1 models at the LHC and future
  hadron colliders}},
  \href{https://doi.org/10.1103/PhysRevD.106.055027}{\emph{Phys. Rev. D}
  {\bfseries 106} (2022) 055027}
  [\href{https://arxiv.org/abs/2203.02520}{{\ttfamily 2203.02520}}].

\bibitem{Oliveira:2022vjo}
V.~Oliveira and C.A.d.~S.~Pires, \emph{{Flavor changing neutral current
  processes and family discrimination in 3-3-1 models}},
  \href{https://doi.org/10.1088/1361-6471/acf1b7}{\emph{J. Phys. G} {\bfseries
  50} (2023) 115002} [\href{https://arxiv.org/abs/2208.00420}{{\ttfamily
  2208.00420}}].

\bibitem{Buras:2012dp}
A.J.~Buras, F.~De~Fazio, J.~Girrbach and M.V.~Carlucci, \emph{{The Anatomy of
  Quark Flavour Observables in 331 Models in the Flavour Precision Era}},
  \href{https://doi.org/10.1007/JHEP02(2013)023}{\emph{JHEP} {\bfseries 02}
  (2013) 023} [\href{https://arxiv.org/abs/1211.1237}{{\ttfamily 1211.1237}}].

\bibitem{Mizukoshi:2010ky}
J.K.~Mizukoshi, C.A.~de~S.~Pires, F.S.~Queiroz and P.S.~Rodrigues~da Silva,
  \emph{{WIMPs in a 3-3-1 model with heavy Sterile neutrinos}},
  \href{https://doi.org/10.1103/PhysRevD.83.065024}{\emph{Phys. Rev. D}
  {\bfseries 83} (2011) 065024}
  [\href{https://arxiv.org/abs/1010.4097}{{\ttfamily 1010.4097}}].

\bibitem{Ruiz-Alvarez:2012nvg}
J.D.~Ruiz-Alvarez, C.A.~de~S.~Pires, F.S.~Queiroz, D.~Restrepo and
  P.S.~Rodrigues~da Silva, \emph{{On the Connection of Gamma-Rays, Dark Matter
  and Higgs Searches at LHC}},
  \href{https://doi.org/10.1103/PhysRevD.86.075011}{\emph{Phys. Rev. D}
  {\bfseries 86} (2012) 075011}
  [\href{https://arxiv.org/abs/1206.5779}{{\ttfamily 1206.5779}}].

\bibitem{Profumo:2013sca}
S.~Profumo and F.S.~Queiroz, \emph{{Constraining the $Z'$ mass in 331 models
  using direct dark matter detection}},
  \href{https://doi.org/10.1140/epjc/s10052-014-2960-x}{\emph{Eur. Phys. J. C}
  {\bfseries 74} (2014) 2960}
  [\href{https://arxiv.org/abs/1307.7802}{{\ttfamily 1307.7802}}].

\bibitem{Sanchez-Vega:2018qje}
B.L.~S\'anchez-Vega, G.~Gambini and C.E.~Alvarez-Salazar, \emph{{Vacuum
  stability conditions of the economical 3-3-1 model from copositivity}},
  \href{https://doi.org/10.1140/epjc/s10052-019-6807-3}{\emph{Eur. Phys. J. C}
  {\bfseries 79} (2019) 299}
  [\href{https://arxiv.org/abs/1811.00585}{{\ttfamily 1811.00585}}].

\bibitem{Abud:1983id}
M.~Abud and G.~Sartori, \emph{{The Geometry of Spontaneous Symmetry Breaking}},
  \href{https://doi.org/10.1016/0003-4916(83)90017-9}{\emph{Annals Phys.}
  {\bfseries 150} (1983) 307}.

\bibitem{Abud:1981tf}
M.~Abud and G.~Sartori, \emph{{The Geometry of Orbit Space and Natural Minima
  of Higgs Potentials}},
  \href{https://doi.org/10.1016/0370-2693(81)90578-5}{\emph{Phys. Lett. B}
  {\bfseries 104} (1981) 147}.

\bibitem{Sartori:2005sh}
G.~Sartori and G.~Valente, \emph{{The radial problem in gauge field theory
  models}}, \href{https://doi.org/10.1016/j.aop.2005.04.016}{\emph{Annals
  Phys.} {\bfseries 319} (2005) 286}.

\bibitem{Talamini:2006wd}
V.~Talamini, \emph{{Affine-P-matrices in orbit spaces and invariant theory}},
  \href{https://doi.org/10.1088/1742-6596/30/1/005}{\emph{J. Phys. Conf. Ser.}
  {\bfseries 30} (2006) 30}
  [\href{https://arxiv.org/abs/hep-th/0607165}{{\ttfamily hep-th/0607165}}].

\bibitem{Kim:1981xu}
J.~Kim, \emph{{General Method for Analyzing Higgs Potentials}},
  \href{https://doi.org/10.1016/0550-3213(82)90040-2}{\emph{Nucl. Phys. B}
  {\bfseries 196} (1982) 285}.

\bibitem{Degee:2012sk}
A.~Degee, I.P.~Ivanov and V.~Keus, \emph{{Geometric minimization of highly
  symmetric potentials}},
  \href{https://doi.org/10.1007/JHEP02(2013)125}{\emph{JHEP} {\bfseries 02}
  (2013) 125} [\href{https://arxiv.org/abs/1211.4989}{{\ttfamily 1211.4989}}].

\bibitem{Heikinheimo:2017nth}
M.~Heikinheimo, K.~Kannike, F.~Lyonnet, M.~Raidal, K.~Tuominen and
  H.~Veerm{\"a}e, \emph{{Vacuum Stability and Perturbativity of SU(3)
  Scalars}}, \href{https://doi.org/10.1007/JHEP10(2017)014}{\emph{JHEP}
  {\bfseries 10} (2017) 014}
  [\href{https://arxiv.org/abs/1707.08980}{{\ttfamily 1707.08980}}].

\bibitem{Guada:2018jek}
V.~Guada, A.~Maiezza and M.~Nemev{\v{s}}ek, \emph{{Multifield Polygonal
  Bounces}}, \href{https://doi.org/10.1103/PhysRevD.99.056020}{\emph{Phys. Rev.
  D} {\bfseries 99} (2019) 056020}
  [\href{https://arxiv.org/abs/1803.02227}{{\ttfamily 1803.02227}}].

\bibitem{Guada:2020xnz}
V.~Guada, M.~Nemev{\v{s}}ek and M.~Pintar, \emph{{FindBounce: Package for
  multi-field bounce actions}},
  \href{https://doi.org/10.1016/j.cpc.2020.107480}{\emph{Comput. Phys. Commun.}
  {\bfseries 256} (2020) 107480}
  [\href{https://arxiv.org/abs/2002.00881}{{\ttfamily 2002.00881}}].

\bibitem{Dias:2005yh}
A.G.~Dias, C.A.~de~S.~Pires and P.S.~Rodrigues~da Silva, \emph{{Naturally light
  right-handed neutrinos in a 3-3-1 model}},
  \href{https://doi.org/10.1016/j.physletb.2005.09.028}{\emph{Phys. Lett. B}
  {\bfseries 628} (2005) 85}
  [\href{https://arxiv.org/abs/hep-ph/0508186}{{\ttfamily hep-ph/0508186}}].

\bibitem{Dong:2006gx}
P.V.~Dong, D.T.~Huong, T.T.~Huong and H.N.~Long, \emph{{Fermion masses in the
  economical 3-3-1 model}},
  \href{https://doi.org/10.1103/PhysRevD.74.053003}{\emph{Phys. Rev. D}
  {\bfseries 74} (2006) 053003}
  [\href{https://arxiv.org/abs/hep-ph/0607291}{{\ttfamily hep-ph/0607291}}].

\bibitem{Montero:2011tg}
J.C.~Montero and B.L.~Sanchez-Vega, \emph{{Natural PQ symmetry in the 3-3-1
  model with a minimal scalar sector}},
  \href{https://doi.org/10.1103/PhysRevD.84.055019}{\emph{Phys. Rev. D}
  {\bfseries 84} (2011) 055019}
  [\href{https://arxiv.org/abs/1102.5374}{{\ttfamily 1102.5374}}].

\bibitem{Cogollo:2014jia}
D.~Cogollo, A.X.~Gonzalez-Morales, F.S.~Queiroz and P.R.~Teles,
  \emph{{Excluding the Light Dark Matter Window of a 331 Model Using LHC and
  Direct Dark Matter Detection Data}},
  \href{https://doi.org/10.1088/1475-7516/2014/11/002}{\emph{JCAP} {\bfseries
  11} (2014) 002} [\href{https://arxiv.org/abs/1402.3271}{{\ttfamily
  1402.3271}}].

\bibitem{Bardeen:1986yb}
W.A.~Bardeen, R.D.~Peccei and T.~Yanagida, \emph{{CONSTRAINTS ON VARIANT AXION
  MODELS}}, \href{https://doi.org/10.1016/0550-3213(87)90003-4}{\emph{Nucl.
  Phys. B} {\bfseries 279} (1987) 401}.

\bibitem{Carvajal:2017gjj}
C.D.R.~Carvajal, B.L.~S{\'a}nchez-Vega and O.~Zapata, \emph{{Linking axionlike
  dark matter to neutrino masses}},
  \href{https://doi.org/10.1103/PhysRevD.96.115035}{\emph{Phys. Rev. D}
  {\bfseries 96} (2017) 115035}
  [\href{https://arxiv.org/abs/1704.08340}{{\ttfamily 1704.08340}}].

\bibitem{LAURENT1995439}
M.~Laurent and S.~Poljak, \emph{On a positive semidefinite relaxation of the
  cut polytope},
  \href{https://doi.org/https://doi.org/10.1016/0024-3795(95)00271-R}{\emph{Linear
  Algebra and its Applications} {\bfseries 223-224} (1995) 439}.

\bibitem{Kannike:2012pe}
K.~Kannike, \emph{{Vacuum Stability Conditions From Copositivity Criteria}},
  \href{https://doi.org/10.1140/epjc/s10052-012-2093-z}{\emph{Eur. Phys. J. C}
  {\bfseries 72} (2012) 2093}
  [\href{https://arxiv.org/abs/1205.3781}{{\ttfamily 1205.3781}}].

\bibitem{Kannike:2016fmd}
K.~Kannike, \emph{{Vacuum Stability of a General Scalar Potential of a Few
  Fields}}, \href{https://doi.org/10.1140/epjc/s10052-016-4160-3}{\emph{Eur.
  Phys. J. C} {\bfseries 76} (2016) 324}
  [\href{https://arxiv.org/abs/1603.02680}{{\ttfamily 1603.02680}}].

\bibitem{ParticleDataGroup:2024cfk}
{\scshape Particle Data Group} collaboration, \emph{{Review of particle
  physics}}, \href{https://doi.org/10.1103/PhysRevD.110.030001}{\emph{Phys.
  Rev. D} {\bfseries 110} (2024) 030001}.

\end{thebibliography}\endgroup

%%%%%%%%%%%%%%%%%%%%%%%%%%%%%%%%%%%%%%%%%%%%%%%%%%%%%%%%%%
\end{document}